\documentclass[journal]{IEEEtran}
\usepackage[table,xcdraw]{xcolor}
\usepackage{makecell}
\usepackage{tabu}

\usepackage[utf8]{inputenc} 
\usepackage[french,english]{babel} 

\usepackage{pifont} 
\usepackage{xcolor}

\usepackage{ifthen} 
\usepackage{amsmath}
\usepackage{amssymb}
\usepackage{amsthm}
\usepackage{siunitx} 
\usepackage{booktabs}
\usepackage[noend]{algpseudocode}
\usepackage[abbreviations]{foreign}

\ifCLASSINFOpdf
\else
\fi

\usepackage{xcolor}
\usepackage{caption}
\usepackage{subcaption}
\usepackage{graphicx}
\usepackage{comment}

\usepackage[utf8]{inputenc}
\usepackage[nopostdot,style=super,nonumberlist,toc,nogroupskip]{glossaries}
\usepackage[table,xcdraw]{xcolor}
\makeglossaries
\setacronymstyle{long-short}
\newacronym{ntma}{NTMA}{Network Traffic Monitoring and Analysis}
\newacronym{iid}{i.i.d}{Identically and independently distributed}
\newacronym{ntc}{NTC}{Network Traffic Classification}
\newacronym{cfs}{CFS}{Correlation based Feature Selection}
\newacronym{rl}{RL}{Reinforcement Learning}
\newacronym{ml}{ML}{Machine Learning}
\newacronym{mape}{MAPE}{Monitor-Analyze-Plan-Execute}
\newacronym{sdn}{SDN}{Software Defined Networking}
\newacronym{rae}{RAE}{Relief Attribute Evaluation}
\newacronym{ooda}{OODA}{Observe-Orient-Decide-Act}
\newacronym{qos}{QoS}{Quality of Service}
\newacronym{qoe}{QoE}{Quality of Experience}
\newacronym{nmo}{NMO}{Network Management and Orchestration}
\newacronym{al}{AL}{Active Learning}
\newacronym{sfem}{SFEM}{Structural Feature Extraction Methodology}
\newacronym{ral}{RAL}{Reinforcement \gls{al}}
\newacronym{ddos}{DDoS}{Distributed Denial of Service}
\newacronym{asvm}{ASVM}{\gls{al} Support Vector Machine}
\newacronym{albl}{ALBL}{\gls{al} by learning}
\newacronym{gan}{GAN}{Generative Adversarial Network}
\newacronym{eer}{EER}{Expected error reduction}
\newacronym{lal}{LAL}{Learning Active Learning}
\newacronym{qbc}{QBC}{Query-By-Committee}
\newacronym{unc}{UNC}{Uncertainty sampling}
\newacronym{gmm}{GMM}{Gaussian Mixture Model}
\newacronym{hat}{HAT}{Hoeffding Adaptive Tree}
\newacronym{arf}{ARF}{Adaptive Random Forest}
\newacronym{sae}{SAE}{stacked autoencoders}
\newacronym{nlp}{NLP}{Natural Language Processing}
\newacronym{dl}{DL}{Deep Learning}
\newacronym{dcgan}{DCGAN}{Deep Convolutional Generative Adversarial Network}
\newacronym{iiot}{IIoT}{Industrial \gls{iot}}
\newacronym{dfr}{DFR}{deep-full-range}
\newacronym{ai}{AI}{Artificial Intelligence}
\newacronym{ids}{IDSs}{Intrusion Detection System}
\newacronym{vpn}{VPN}{Virtual Private Network}
\newacronym{dpi}{DPI}{Deep Packet Inspection}
\newacronym{m2m}{M2M}{Machine-to-Machine}
\newacronym{iot}{IoT}{Internet of Things}
\newacronym{vae}{VAE}{Variational Autoencoder}
\newacronym{cnn}{CNN}{Convolutional Neural Network}

\newacronym{sdae}{SDAE}{Stacked Denoising Autoencoder}
\newacronym{lstm}{LSTM}{Long Short-Term Memory}
\newacronym{wsn}{WSNs}{Wireless Sensor Networks}
\newacronym{p2p}{P2P}{Peer-to-peer}
\newacronym{tls}{TLS}{Transport Layer Security}
\newacronym{mlp}{MLP}{Multi-layer Perceptron}
\newacronym{svm}{SVM}{Support Vector Machine}
\newacronym{svdd}{SVDD}{Support Vector Data Dscription}
\usepackage{amsmath}
 \usepackage{setspace}
\makeatletter

\begin{document}
%
\font\myfont=cmr12 at 12pt
\title{ Active Learning for Network Traffic Classification: A Technical \textcolor{black}{Study}}
%
%
%

\author{ Amin Shahraki,
        Mahmoud Abbasi, Amir Taherkordi 
        and~Anca Delia Jurcut
\thanks{Amin Shahraki is with School of Computer Science, University College Dublin, Ireland. Corresponding author e-mail: (am.shahraki@ieee.org)} 

\thanks{Mahmoud Abbasi was with Department of Computer Sciences, Islamic Azad University, Mashhad, Iran, email: mahmoud.abbasi@ieee.org}%
\thanks{Amir Taherkordi is with the Department
of Informatics (IFI), University of Oslo, Norway. email: amirhost@ifi.uio.no}
\thanks{Anca Delia Jurcut is with Department of Computer Sciences, University College Dublin, Dublin, Ireland, email: anca.jurcut@ucd.ie}
}

\markboth{ IEEE Transactions on Cognitive Communications and Networking}%
{Shahraki \MakeLowercase{\textit{et al.}}:}

%



\maketitle
\textit{\textbf{\textcolor{blue}{[Note: This work has been submitted to the IEEE Transactions on Cognitive  Communications  and  Networking  journal for possible publication. Copyright may be transferred without notice, after which this version may no longer be accessible]}}}

\begin{abstract}
\textcolor{black}{Network Traffic Classification (NTC) has become an \textcolor{black}{important feature in various network} management operations, \eg, Quality of Service (QoS) provisioning and security services. Machine Learning (ML) algorithms as a \textcolor{black}{popular approach for NTC can promise reasonable accuracy in classification and deal with} encrypted traffic. However, ML-based NTC techniques suffer from the shortage of labeled traffic data which is the case in many real-world applications. This study investigates the applicability of an active form of ML, called Active Learning (AL), in NTC.
AL reduces the need for a large number of labeled examples by actively choosing the instances that should be labeled. The study first provides an overview of NTC and its fundamental challenges along with surveying the literature on ML-based NTC methods. Then, it introduces the concepts of AL, discusses it in the context of NTC, and review the literature in this field. Further, challenges and open issues in \textcolor{black}{AL-based classification of network traffic are discussed. Moreover,} as a technical survey, some experiments are conducted to show the broad applicability of AL in NTC. The simulation results show that AL can achieve high accuracy with a small amount of data. 
}
\end{abstract}

\begin{IEEEkeywords}
Survey, Network Traffic Classification, Active Learning, Machine Learning, NTMA
\end{IEEEkeywords}

%
\IEEEpeerreviewmaketitle

\section{Introduction} \label{introduction}
%
%
%
%


\label{sec:intro}
During the last decades, emerging new networking paradigms, \textcolor{black}{such as \gls{iot}, have introduced} various network management challenges.
\textcolor{black}{Given the proliferation of \gls{iot} devices and the distinguishing characteristics of IoT traffic, such as heterogeneity, spatio-temporal dependencies, dominating uplink traffic, and low duty-cycle traffic patterns, network management and monitoring has become challenging.}
Gaining deep insight into such complex networks for performance evaluation and network planning purposes is not a trivial task \textcolor{black}{with respect to} processing time, human effort, and computational overhead. Understanding network traffic behavior plays a vital role in a wide variety of network management aspects, \eg, fault management, accounting, security, and network performance management~\cite{wang2013internet}. Some general approaches have been introduced to analyze the behavior of networks and maintain \textcolor{black}{their performance, such as \gls{mape}, and \gls{ooda} \cite{ayoubi2018machine}}.

In networking, the process of analyzing the network traffic behavior is mainly known as \gls{ntma}~\cite{casas2016big}. 
\gls{ntma} has attracted much interest in recent years and become an important research topic in the field of communication systems and networks \cite{8789667}. The importance of \gls{ntma} lies in the properties and challenges of modern networking, \eg, heterogeneity, complexity, and dynamicity, resulting in instability in data transmission \cite{stoyanova2020survey}.  
\gls{ntma} is an essential approach to measure the performance of applications and services, and to discover network inefficiencies. Indeed, \gls{ntma} allows us to shed light on the functioning of communication systems and to deal with unexpected events, especially in complex and large-scale networks, such as the Internet. 

\textcolor{black}{\textcolor{black}{\gls{ntma} applications are generally categorized into eight groups}, including \textit{\gls{ntc}}, \textit{traffic prediction}, \textit{fault management}, \textit{network security}, \textit{traffic routing}, \textit{congestion control}, \textit{resource management}, and \textit{\gls{qos} and \gls{qoe} management}} \cite{boutaba2018comprehensive}. In this study, we focus on \gls{ntc} as an important and open issue in \gls{ntma}. \gls{ntc} refers to techniques for categorizing network traffic into different classes based on their properties.
The classification of network traffic is highly beneficial in various network services from \gls{qos} (\eg, traffic policing and shaping) and pricing to malware and intrusion detection \cite{tahaei2020rise}.
\textcolor{black}{
\gls{ntc} provides detailed knowledge on network traffic,
which is very useful for those who investigate the changes in traffic characteristics and long-term requirements of networks \cite{4738466}, \eg \gls{nmo} tools, and performance management models.
}

\gls{ntc} techniques can be broadly grouped into three categories: \textcolor{black}{ \textit{port-based}, \textit{payload-based}, and \textit{flow-based} methods \cite{wang2019survey}.
Port-based techniques associate a standard port number to a service or application, while payload-based methods carefully inspect the content of the captured packets to classify them. Last but not least, flow-based techniques utilize the network traffic flow characteristics (\eg, round-trip time and inter-arrival times) to associate produced traffic to the related sources. }\textcolor{black}{The two latter methods cannot be used in some network types (\eg \gls{vpn}), or violate the privacy of users by accessing their personal data.}
Flow-based techniques are the most common techniques for \gls{ntc} as instead of inspecting all packets passing through a given link, they examine network traffic flows or an aggregated form of the network header packets information. As a result, the volume of data needed to be examined will be reduced, and the encrypted traffic will no longer be a problem. Flow-based techniques assume that each application's traffic has almost unique statistical or time-series features that \textcolor{black}{can be utilized by classifiers} to categorize both encrypted and regular traffics.

\textcolor{black}{In flow-based methods, the traffic classifier may leverage \gls{ml} algorithms to automate the classification process, \textcolor{black}{ discover different traffic patterns produced by devices, and classify} encrypted traffic.}
 \textcolor{black}{Although \gls{ml} algorithms are powerful techniques to classify network traffic flows \cite{lopez2020iot, 8026581}, the accuracy of learning-based approaches is limited by their need for a \textit{massive number of labeled} instances. As the authors in \cite{alsheikh2016mobile} mentioned, most of the real-world application data is semi-labeled or unlabeled data. Moreover, the data labeling process for ML tasks can be challenging in terms of human effort and cost \cite{tu2020better}.}



Fortunately, \gls{al}, as a sub-field of \gls{ml}, is a promising approach to deal with \textit{the need for a huge amount of labeled instances}. \gls{al} aims to reduce the need for labeled examples by intelligently querying the labels during training. The query goes for the examples that the AL algorithm believes will help build the best model \cite{settles1648active}. 
\textcolor{black}{Therefore, based on the aforementioned challenges, \gls{al} can be considered as an appropriate and efficient technique for flow-based \gls{ntc}. Providing a thorough study on the usefulness of \gls{al} in \gls{ntc} and reviewing the state-of-the-art techniques in this field can significantly help the network research community in better adoption of AL for classification of network traffic in various domains. To the best of our knowledge, this is the first and only study that technically reviews the efficiency and importance of \gls{al} for \gls{ntc} along with surveying the literature in this field.}
In this paper, we study the \gls{ntc} techniques and discuss \gls{al} as a useful approach in this field. The main contributions of our work are summarized as follows: 
\begin{itemize}
\item Discussing \gls{ntc} techniques and their correlations with \gls{ml} techniques
\item Reviewing existing work in \gls{al}-based \gls{ntc}
\item Empirical evaluation of the performance of \gls{al} for \gls{ntc} purposes
\item Discussing the challenges, and future directions in using \gls{al} for \gls{ntc} 

\end{itemize}
\textcolor{black}{The rest of this paper is structured as follows: \textcolor{black}{In Section \ref{relatedworks}, we review existing survey works on traffic classification techniques.}
In Section \ref{sec:ntc}, we provide an overview of the \gls{ntc} problem and the use of \gls{ml} techniques. Then, we devote Section \ref{sec:al} to discussing the fundamental elements of \gls{al} and query strategies. Next, in Section \ref{sec:active NTC}, we discuss \textcolor{black}{the advantages of using \gls{al} for \gls{ntc} purposes and carry out a literature review on this topic.} In Section \ref{sec:performance}, we evaluate the performance of \gls{al} in \gls{ntc}. In Section \ref{sec:challenges}, we \textcolor{black}{discuss the challenges and future directions in} using \gls{al} for \gls{ntc}, and finally we conclude the paper in Section \ref{sec:conclusion}.}
\textcolor{black}{\printglossary[title=List of abbreviations,nonumberlist]}

\section{Related Survey Articles} 
\label{relatedworks}
\textcolor{black}{
There exist several literature studies reviewing the use of \gls{ml} techniques in communication systems and wireless networks, \eg \cite{sun2019application,chen2019artificial}. There are also some surveys that focus on specific \gls{ml} techniques, \eg \gls{dl} \cite{mao2018deep} and \gls{rl} \cite{yau2012reinforcement} , or specific types of networking, \eg \gls{sdn} \cite{xie2018survey} and optical networks \cite{gu2020machine}. Moreover, some survey works compare, evaluate or review different techniques, \eg \gls{ml}-based techniques, heuristic models and statistical-based techniques for \gls{ntc} \eg \cite{singh2015performance}. Considering the volume of survey literature in this field, in this section, we focus only on surveys that review \gls{ntc} or the use of various \gls{ml} techniques in \gls{ntc}.}

\textcolor{black}{\begin{itemize}
 \item \textbf{\textit{General literature reviews on \gls{ntc}:}} In \cite{dainotti2012issues}, Dainotti et al. reviewed the issues and future research directions of \gls{ntc}, especially in case of applicability, reliability and privacy. They outlined the research and policy future directions of \gls{ntc}, \eg validating the \gls{ntc} models, \textcolor{black}{effect of network speed} in \gls{ntc} and \gls{ntc} tools. In \cite{finsterbusch2013survey}, Finsterbusch et al. reviewed the payload-based \gls{ntc} based on \gls{dpi}. They also practically analysed the most significant open-source \gls{dpi} modules to show their performance in terms of accuracy and requirements. Additionally, they provided a guideline on how to design and implement \gls{dpi}-based \gls{ntc} modules. In \cite{velan2015survey}, Velan et al. studied \gls{ntc} models for encrypted network traffics to measure the traffic and improve the security, \eg detecting anomalies. They have reviewed different types of encrypted traffics and how payload-based and feature-based \gls{ntc} techniques can classify encrypted network traffics. Zhao et al. \cite{tahaei2020rise} reviewed the use of \gls{ntc} in \gls{iot} and \gls{m2m} networks. They reviewed the current \gls{ntc} within the IoT context based on the differences between IoT and non-IoT network traffics. By reviewing the literature, the authors showed that in IoT research area, most of \gls{ntc} techniques are proposed to solve security challenges. The authors in \cite{zhao2021network} reviewed the \gls{ntc} techniques, \ie statistics-based classification, correlation-based classification, behavior-based classification, payload-based classification, and port-based classification. They also quantified classification granularity based on four levels, \ie application type layer, protocol layer, application layer and service layer. Last but not least, they classified network traffic features and the existing public datasets that are commonly used in the proposed \gls{ntc} techniques.
 \item \textbf{\textit{Literature reviews on the use of \gls{ml} in \gls{ntc}:}} As one of the earliest study in the use of \gls{ml} in \gls{ntc}, Nguyen et al. \cite{nguyen2008survey} reviewed the literature between the years 2004 to 2007. They studied how \gls{ml} models can be employed for \gls{ntc} in IP networks, \eg clustering approaches, supervised learning approaches and hybrid approaches. They also reviewed the literature that compares ML techniques or non-ML techniques for \gls{ntc}. They mentioned that offline analysis models, \eg AutoClass, Decision Tree and Naive Bayes can achieve a high accuracy for about 99\%. They also outlined some critical operational requirements for real-time \glspl{ntc} models compared to offline models. In \cite{singh2015performance}, \textcolor{black}{Singh evaluated} the unsupervised \gls{ml} techniques including \textit{K}-means and Expectation Maximization algorithm for \gls{ntc}. The results show that the accuracy of \textit{K}-Means is better than Expectation Maximization algorithm. In \cite{perera2017comparison}, Perera et al. compared six \gls{ml} algorithms including Naive Bayes, Bayes Net, Naive Bayes Tree, Random Forest, Decision Tree and Multiplayer Perceptron along with two feature extraction techniques, \ie \gls{cfs} and \gls{rae}. Their results show that Decision Tree and Random Forest have better performance compared to other techniques. In \cite{gomez2017ensemble}, Gomez et al. compared seven ensemble \gls{ml} techniques including OneVsRest, OneVsOne, Error-Correcting Output-code, Adaboost classifier, Bagging algorithm, Random Forest and Extremely Randomized Trees which are all based on decision trees in \gls{ntc}. They compared them in case of model accuracy, latency and byte accuracy. In \cite{pacheco2018towards}, Pacheco et al. comprehensively surveyed the use of \gls{ml} techniques in \gls{ntc} for different cases, \eg encrypted network traffic. By understanding the challenges of using \gls{ml} techniques in \gls{ntc}, they studied the reliable label assignment, dynamic feature selection, integrating the meta-learning processes. \textcolor{black}{They considered these solutions to solve several issues, including imbalance network data, dynamicity of networks, and online strategies for re-training the ML models.}
\end{itemize}}
 
In Table~\ref{tab:existingSurveys}, a summary of the surveys above is provided based on \textcolor{black}{their vision of \gls{ntc},} the reviewed solutions, network type and practical evaluation of studied solutions.
As indicated in the table, our survey is for flow-based \gls{ntc} for the use in Internet communications and \textcolor{black}{specifically considers} \gls{al} as one of the most important \gls{ml}-based solution.
To the best of our knowledge, our study is one of the rare literature surveys that evaluates such specific \gls{ml} solutions for \gls{ntc} as most of existing surveys consider general \gls{ml} models, \eg supervised learning solutions for \gls{ntc}. 
Studying AL-based solutions makes our work different from all existing survey works.

\begin{table*}[h]
\scriptsize
\centering
\caption{\textcolor{black}{An overview of existing literature surveys on \gls{ntc} and ML.}}
\label{tab:existingSurveys}
\begin{tabular}{|l|l|p{3cm}|l|l|l|}
\hline
Study & Year & NTC vision & Reviewed Solution(s) & Type of network & \makecell{Practical\\ Evaluation} \\ \hline
 \cite{nguyen2008survey}& 2008& Analysing Statistical traffic Characteristics & ML Solutions & IP networks & No \\ \hline
 \cite{dainotti2012issues}&2012& General \gls{ntc} & Not Specified& TCP Networks & No \\ \hline
 \cite{finsterbusch2013survey} &2014& Payload-Based \gls{ntc} techniques & \gls{dpi}-based techniques & Internet &Yes \\ \hline
 \cite{singh2015performance} &2015& Comparative Study & Comparing unsupervised \gls{ml} techniques & Internet & Yes \\ \hline
 \cite{velan2015survey} &2015& Analysing encrypted network traffic by payload-based and feature-based \gls{ntc} technique & ML techniques and hybrid techniques & Not Specified & No \\ \hline
 \cite{perera2017comparison} &2017& Comparative Study & Comparing six \gls{ml} Solutions & Communication Networks & Yes \\ \hline

 \cite{gomez2017ensemble} & 2017& Comparative Study & Comparing Decision-tree based ensemble techniques & Internet & Yes \\ \hline
 \cite{pacheco2018towards} & 2018& \gls{ml}-based \gls{ntc} & Most existing \gls{ml} solutions & IP Networks & No \\ \hline
 
 \cite{tahaei2020rise} & 2020 &\gls{ntc} for \gls{m2m} network traffic& Generic solutions & \gls{iot} &No \\ \hline
 \cite{zhao2021network} &2021& Reviewing various types of \gls{ntc} models & Most existing \gls{ml} solutions & Internet &No \\ \hline
  \textcolor{black}{Our Study}&2021& Flow-based \gls{ntc} & Active Learning& Internet&Yes \\ \hline
 
\end{tabular}
\end{table*}

\section{Overview on \gls{ntc} and \gls{ml}}
\label{sec:ntc}
\textcolor{black}{
In \gls{ntc}, one should clarify the goals of classification based on the intended use, such as for accounting purposes, malware detection, intrusion detection, providing \gls{qos}, and identifying types of applications based on the network traffic (\eg, \gls{vpn} and nonVPN traffics or Tor and nonTor traffics). Indeed, there are different factors that one can use to categorize network traffic, including applications (\eg, Facebook and Hangouts), protocols (\eg, HTTP and BitTorrent), traffic types (\eg, Web Browsing and Chat), browsers (\eg, Firefox and Chrome), operating systems, and websites. Therefore, the purpose is to determine the label of each network flow truly, \eg, browsing, interactive, and video stream. \gls{ntc} can be further categorized into \textit{online} and \textit{offline} classification. In online \gls{ntc}, the input traffic needs to be classified in a real-time or near real-time manner (\eg, \gls{qos} provisioning). On the other hand, offline classification is appropriate for applications such as anomaly detection and billing systems.}
Despite their importance, existing \gls{ntc} techniques suffer from general networking challenges as listed below:

\begin{itemize}
\item \textcolor{black}{While the literature on traffic classification is mature to adapt to old-fashioned networking paradigms, \eg, legacy cellular systems,
the dramatic growth and evolution of online applications and services have made traffic classification a non-trivial task.
Due to the traffic characteristics of modern networks, \eg, being large-scale, heterogeneity, multimodal data, and big data, emerging \gls{ntc} methods must meet strict requirements in terms of system performance, accuracy, and robustness.
For example, the vast amount of raw data generated by \gls{iot} and cellular devices pose severe challenges to \gls{ml}-based \gls{ntc} methods as they need clean and pre-processed data for training purposes.}

\item \textcolor{black}{\gls{ntc} is a multi-factor procedure in which an automated program categorizes the network traffic based on the network traffic features, \eg, types of network protocols, applications, hosts, etc. As a challenge, \gls{ntc} techniques need to select the \textit{best features} to classify the network traffic with high accuracy, while each of them can be efficient or inefficient from one network to another network.
In other words, \textit{feature engineering} is a challenge when it comes to using classical \gls{ml} for traffic classification.}

\item \textcolor{black}{The recent increase of encrypted network traffic and protocol encapsulation methods limit the effectiveness of many traffic classification techniques since the packet inspection techniques are unable to extract network management information from network traffics.
For example, a significant portion of the Internet traffic is associated with \gls{p2p} applications. However, classification of \gls{p2p} traffic is a difficult task \cite{tahaei2020rise} as many \gls{p2p} applications, such as online video and \gls{p2p} downloading, use encryption and obfuscation protocols to remove the limitations posed by Internet service providers.}
\end{itemize}

\begin{figure}[h]
 \centering
 \includegraphics[width=9cm]{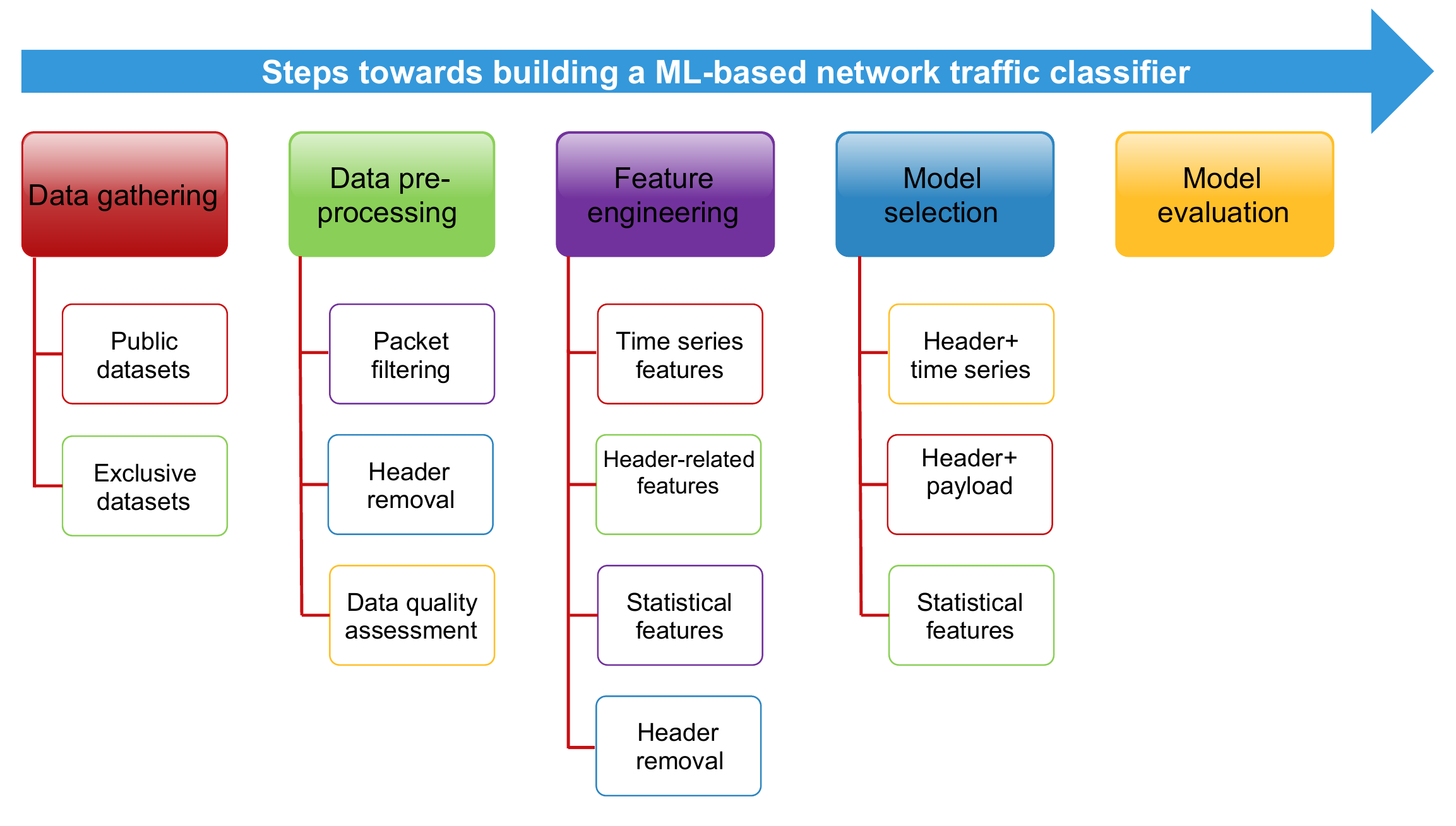}
 \caption{The main steps in building a network traffic classifier.}
 \label{fig:steps}
\end{figure}

\textcolor{black}{To overcome the above challenges, various techniques have been introduced, \eg graphical techniques, statistical methods and \gls{ml}-based methods \cite{velan2015survey}. In the scope of \gls{ml}, various solutions for port-based, payload-based, and flow-based have been proposed as the most promising solutions for \gls{ntc} \cite{moore2005internet} \cite{yuan2010svm}.}
\textcolor{black}{Multiple steps are needed for building a ML-based network traffic classifier as presented in \cite{rezaei2019deep}. Figure \ref{fig:steps} shows a graphical description of all steps.}
\textcolor{black}{In the rest of this section, we discuss each individual step.}

\subsection{Data gathering}
Since ML algorithms learn to classify the data based on sample datasets, representative data must be collected as the data gathering step. While a few publicly available network traffic datasets have been released, using these to train a traffic classification model can be difficult \cite{ring2019survey}. In addition, since the behavior of the network traffic is different from one network to another one, it is highly recommended to train the \gls{ml} algorithm for the target network \cite{ayoubi2018machine}. Additionally, the number of network traffic classes can be high, and it is rather impractical to consider all classes in one public dataset.
Furthermore, there are a variety of data gathering and labeling techniques
that lead to different feature sets. Hence, in real-world applications, the goal is to use datasets that are tailored to the intended use of \gls{ntc}, mainly gathered from the target network.

\subsection{Data pre-processing}
After gathering, the data must be pre-processed such that it is represented in a form that \textcolor{black}{the target ML algorithms can discover different patterns}. In traffic classification, \textit{header data} and \textit{payload} are two major data structures. These structures often need to be pre-processed because they contain irrelevant or redundant information, such as network management data, which is not needed for traffic classification, \eg, source and destination IP addresses, and protocol information.
Moreover, changes in the distribution of packet-level features can occur in real-world environments because of unexpected events like the re-transmission of packets. In short, performing some pre-processing steps such as packet filtering, elimination of noisy samples, header removal, and data quality assessment is needed to ease the learning process for the ML algorithms \cite{kamiran2012data}.

\textcolor{black}{
\subsection{Feature engineering}
Conventional classification solutions, \eg ML- and statistical-based techniques, need to go through a feature engineering procedure, in which domain knowledge is used to extract features or patterns from the raw data \cite{zheng2018feature, finsterbusch2013survey}. 
Feature engineering is a crucial step in ML-based \gls{ntc} methods because of the fact that choosing appropriate features can ease the difficulties of the modelling phase, and vice versa \cite{dong2018feature}. 
\textcolor{black}{It is worth mentioning that considering privacy, the risk associated with feature engineering and representation procedures is also crucially important, especially in the payload feature-based techniques.} Indeed, there are some legal restrictions on using payload-based methods in many environments or recognizing all communication protocols. This is mainly due to the user’s privacy policies, as such methods inspect the content of the network packets \cite{sicker2007legal}. 
}

Generally, there are four major types of input features for NTC: 
\begin{itemize}
 \item \textbf{Time series:} Considering time series related features, one can refer to maximum packet inter-arrival time, maximum number of bytes in packet, and inter-packet timings. According to \cite{lopez2017network}, the length of time series (or the number of packets within a flow) has a visible effect on classification accuracy and computational overhead. Specifically, increasing the number of considered packets can improve the classification performance but at the cost of higher computational overhead. 
 In \cite{lopez2017network}, only the first 20 traffic packets in a flow are used for the experiments. The authors in \cite{rezaei2018achieve} use the time-series features of packets, \textcolor{black}{\eg, source and destination ports, payload size, and \textcolor{black}{TCP window size (bytes)} as input for a semi-supervised model to perform traffic classification related to the five Google services, including Hangout Chat, Hangout Voice Call, YouTube, File transfer, and Google play music.
 The simulation result shows excellent accuracy, despite using a limited number of labeled data samples. This is mainly because they conducted a pre-training step on the entire unlabeled network flows in order to learn statistical features, and then they re-trained the model using a small labeled dataset for fine-tuning.}
 
 \item \textbf{Header:} The header of a network packet contains information related to different layers (\eg the network layer). Features such as port number and protocol number are widely used as informative features in traffic classification tasks. However, some modern NTC techniques, especially \gls{dl}-based, \textcolor{black}{accept entire packets as the input feature.
 For example, in \cite{kim2018tor} the authors used hexadecimal raw packet header and convolutional networks to classify Tor/non-Tor traffic. To this end, they utilized TCP/IP headers, especially the first 54 bytes of packets, because TCP is associated with around 90\% of all the Internet traffic.}
 
 \item \textbf{Payload:} NTC techniques can also use layer-related information above the transport layer to classify network traffic. As a prime example, in \cite{li2011quick} the authors utilize BitTorrent handshake packets on layer 4 to classify the \textcolor{black}{BitTorrent traffic. 
BT generates the highest amount of \gls{p2p} traffic.} Moreover, some works use packets related to the \gls{tls} handshake process to identify HTTPS services \cite{shbair2016multi}.
 
 \item \textbf{Statistical features:} The statistical features of network flows, such as minimum inter-arrival time and size of the IP packets can be used for \gls{ntc} \cite{crotti2007traffic}. The main idea behind using statistical features is that the statistical features of network flows generated by different services or applications are almost unique.
 \textcolor{black}{Nevertheless, a big challenge with the methods that use statistical features is that they are not suitable for online classification.}
 This is mainly due to the fact that a classifier needs to monitor the entire or significant part of a network flow in order to extract statistical features.
\end{itemize}

\subsection{Model selection}
Another step towards building a traffic classifier is selecting the right \gls{ml} model. In the context of ML, choosing a model can carry different meanings, such as the selection of hyper-parameters and parameters, as well as algorithm selection. Given \gls{ntc}, several factors can be involved in the selection of the classification model \textcolor{black}{(\eg, model performance, available resources, model complexity, and feature selection).
One of the most significant factors is feature selection.
This is due to the fact that there is a direct correlation between features and input dimensions of the model, and consequently the computational and memory complexities of the model, which are crucial factors in \gls{ntc}.}
This implies that the dimensions and structure of input data for training purposes should be optimized.
\textcolor{black}{Moreover, the selected features directly affect the performance of the final learning task (\eg, classification and regression) and the dimensions of the input data for training. Hence, one should consider the right number of informative features.} 
\textcolor{black}{In the context of traffic classification, it may be not sufficient to consider the model performance as the only factor for model selection. Thus, one can also consider other criteria, such as training time and model explainability.}


\subsection{Model Evaluation}
Finally, the evaluation of the selected model is the final step in building a network traffic classifier. In this step, the performance of the ML model on unseen data is measured. The ML model should be able to give accurate predictions to be useful for the given task. However, the accuracy is not the only evaluation metric for a classification task, and other metrics such as confusion matrix, F1 score, recall, etc. should be considered.
\textcolor{black}{NTC is a classification task, and we use the same metrics to evaluate the performance of the proposed model.}

 \subsection{Existing Work}
\textcolor{black}{Recently, several ML techniques have been proposed for network traffic classification.
In this subsection, we categorize existing work in the literature based on the goals of network traffic classification}, including identifying applications (also called apps), cyber security purposes, fault detection, website fingerprinting, user activities identification, and operating systems identification. We discuss these goals in more details in the sequel.
 
 \begin{itemize}
 \item\textcolor{black}{\textbf{Mobile apps identification:} This goal refers to analyzing and finally identifying the network traffic related to a particular mobile app. Given the ever-increasing number of mobile apps, network administrators and telecommunications companies are actively looking for rigorous methods to secure their infrastructure. Apps identification based on analyzing the network traffic of mobile apps can assist network administrators with resource management and planning, and app-specific policy establishment (\eg, security policy establishment and access management for a specific app). Furthermore, the identification of apps can help protect smartphone platforms (\eg, Android) against emerging security threats and uncover sensitive apps. Moreover, by app identification, it is possible to forbid the use of some particular apps (\eg, Google+ and Instagram) in an enterprise network \cite{ajaeiya2018mobile}.
 }
 \textcolor{black}{Several papers have been published on app identification.
 Ajaeiya \textit{et al.} in \cite{ajaeiya2018mobile} present a framework for the classification of Android apps. The proposed framework identifies apps traffic from a network viewpoint without adding any overhead on users' mobile phones. Moreover, the authors provide a pre-processing method for traffic flows to extract the most informative features for \gls{ml}-based techniques. The work in \cite{li2017traffic} leverages \gls{vae} for the identification of mobile apps. The authors claimed that their method is able to label a massive number of instances and extract \textcolor{black}{the features in} mobile apps traffic automatically. To this end, the authors first transform the mobile apps traffic to meaningful images, and then use \gls{vae} as a classifier. Similar work was carried out by Wang \textit{et al.} in \cite{wang2020real}, in which the authors design three \gls{dl}-based models, including \gls{sdae}, 1D \gls{cnn}, and \gls{lstm} for mobile apps identifications. The authors in \cite{aceto2017traffic} provide a multi-classification scheme for the classification of mobile apps traffic. More specifically, they combine several mobile traffic classifiers' decisions (knowledge) to classify their traffic samples.}
 
 \item \textcolor{black}{\textbf{Cybersecurity purposes:} One of the main goals of traffic classification is detecting security breaches in communication systems, \eg intrusion detection, malware detection, anomaly detection, and worm detection. Cybersecurity tools/techniques (\eg, intrusion detection systems) aim to defend communication systems from internal/external threats. 
 Traffic classification methods can be used to assess network traffic \textcolor{black}{ behavior through detecting malicious traffic flow/link, and then prevent attacks.
 A large body of work in the literature has focused on ML-based malware and intrusion detection.} The authors in \cite{vinayakumar2019deep} propose an intrusion detection approach based on deep neural networks and compare the performance of \gls{dl} with classical ML classifiers, demonstrating the superiority of DL models. Similarly, in \cite{shone2018deep}, Shone \textit{et al.} propose a non-symmetric deep auto-encoder-based learning solution for intrusion detection. The auto-encoder network has been used for learning features in an unsupervised manner. Then, they employ a stacked non-symmetric auto-encoder as a traffic classifier. In \cite{nguyen2019diot}, Nguyen \textit{et al.} propose a federated self-learning method to detect anomalies in \gls{iot} systems.
 Similarly, in \cite{rey2021federated}, Rey \textit{et al.} utilize federated learning for malware detection in IoT devices through one supervised model (based on \gls{mlp}) and one unsupervised model (based on autoencoder).
 To evaluate the framework, they use N-BaIoT dataset, which models the traffic of IoT systems impacted by malware.
 In \cite{mclaughlin2017deep}, McLaughlin \textit{et al.} present a DL-based method for Android malware detection using the raw opcode sequence as the input of a CNN model which can automatically learn the features of malware instances.
 The authors claimed that the proposed method has a more straightforward training pipeline than the previously proposed works (\eg, n-gram-based malware detection). Huang \textit{et al.} \cite{huang2021network} combine the unsupervised spatiotemporal encoder with LSTM to detect abnormal network traffic. The spatial feature of network traffic data was extracted in the first stage by the spatiotemporal model. Then, the obtained features are used to train another LSTM layer for the classification purpose. NSL-KDD dataset was used for the evaluation of the model. Based on the experimental results, using the proposed DL model, the efficiency of intrusion detection is significantly high compared to the traditional techniques.}
 
 \item \textcolor{black}{\textbf{Fault detection:} Fault detection is part of a more extensive network management process, called fault management. Fault management points to a set of processes to detect, isolate, and then correct unusual situations of a network. Failure occurs when a system (\eg, an IoT network) cannot adequately provide a service, where a fault is the source cause of a failure. Fault management, especially fault detection, play an essential role in today’s network management (\eg, QoS provisioning). \textcolor{black}{ Hence, many works have been conducted to improve the fault management process.} In \cite{huang2020machine}, Huang \textit{et al.} survey fault detection techniques in IoT networks and introduce a fault-detection framework for Self-Driving Network (SelfDN)-enabled IoT. Moreover, the authors propose an algorithm called Gaussian Bernoulli restricted Boltzmann machines auto-encoder to change the fault-detection into a classification task. The simulation result demonstrates the superiority of the proposed method to other adopted methods, such as linear discriminant analysis and SVM. In \cite{mulvey2018cell}, the authors focus on the problem of cell coverage degradation detection through a deep neural network. They propose a deep recurrent model for diagnosing cell radio performance deterioration and complete cell outages in a mobile phone network. In \cite{noshad2019fault}, Noshad \textit{et al.} adopt the Random Forest classifier for fault detection in \gls{wsn}. They use a dataset with six types of faults at the sensor levels for performance evaluation, such as data loss, offset, and out-of-bounds. Moreover, they compare the performance of the proposed method with other well-known techniques, \eg, \gls{mlp}, CNN, and probabilistic neural networks.}
 
 \item \textcolor{black}{\textbf{Website fingerprinting:} It refers to methods for identifying and collecting data about websites visited by a mobile device, which is essential for the advertising industry, identifying the characteristics of attacks (\eg, botnets and sniffing) and protecting users' privacy. Website fingerprinting can help recognition of fraudsters and other unusual activities. Moreover, website fingerprinting can be considered as a type of traffic analysis attack that allows eavesdroppers to get information on the victim’s activities. Given the importance of website fingerprinting, there is a large body of literature on this topic.
 In \cite{rahman2020mockingbird}, Rahman \textit{et al.} leverage the idea of adversarial ML to defend users against website fingerprinting attackers. The authors propose a method to generate adversarial examples to decline the accuracy of the attacks that use learning-based techniques for robust traffic classification. The simulation results show that the proposed method can decline the accuracy of the state-of-the-art attack by half. The work in \cite{attarian2019adawfpa} focuses on the concept drift problem in static website fingerprinting attacks for the Tor network. The authors refer to the fact that it is costly to update static attacks in dataset updating and retrain the model. Hence, they introduce AdaWFPA, an adaptive online website fingerprinting attack that leverages adaptive stream mining techniques. 
 Luo \textit{et al.} in \cite{luo2021rbp} propose Random Bidirectional Padding (RBP), a website fingerprinting obfuscation technique against intelligent fingerprinting attacks.
 \textcolor{black}{It uses time sampling and random bidirectional packets padding to change the inter-arrival time characteristics in the traffic flow, and consequently, to identify more complex patterns in network packets.}}
 
 \item \textcolor{black}{\textbf{User activities identification:} Such traffic analysis can be used to obtain exciting pieces of information about a specific action that a mobile subscriber carries out on his/her device (\eg, posting a video on Twitter). The identification of the user activities may also be made to get information about a specific activity, such as the length of a message sent by a user within a particular chat application. User activity identification can be utilized by adversaries/researchers to reveal the identity behind an unknown user, \eg, in a social media, that prefers to remain anonymous. This can be done by behavioral profiling for the users of a network, which is helpful for identifying reconnaissance within the network. Moreover, such traffic analysis offers a possibility to characterize the users' habits in a network, \eg, chatting with friends in the morning and watching the video stream in the evening. \textcolor{black}{The user’s behavior information can be employed next time to detect the user presence in the network. In \cite{7265055}, Conti \textit{et al.} use ML techniques (\ie, Dynamic Time Warping (DTW), hierarchical clustering, and Random Forest) for analyzing Android encrypted network traffic, and consequently, to identify user actions (\eg, email actions, including sending email, replying, and Facebook actions). The authors in \cite{grolman2018transfer} leverage transfer learning to analyze encrypted mobile traffic to deal with the problem of diversity of app releases, mobile operating systems, and model of devices, and identify user actions. 
 The work in \cite{wu2020instagram} focuses on the identification of the Instagram user behavior. Unlike previous works that used the statistical features of encrypted traffic, this work provides a new technique based on maximum entropy to obtain the more stable traffic features. Hou \textit{et al.} in \cite{hou2018classifying} categorize user activities of the WeChat application by performing a detailed analysis on the encryption protocol of this application, called MMTLS, to find the typical user activities of the application (\eg, advertisement click and browsing moments).} Then, they adopt different learning algorithms, such as Naive Bayes, Random forest, and Logistic Regression, to classify these activities.}

 \item \textcolor{black}{\textbf{Operating systems identification:} This refers to identifying the operating system installed on a mobile device by analyzing its generated traffic. Adversaries can use operating systems identification to launch more serious attacks against a specific mobile operating system. Moreover, it is desirable to use this analysis to investigate the popularity of the mobile operating systems (\eg, Android and iOS) among users.
 Hagos \textit{et al.} in \cite{hagos2020machine} introduce a learning-based technique for passive operating systems fingerprinting. They use classical ML (\ie, \gls{svm}, Random Forest, k-nearest neighbors, and Naive Bayes) and DL algorithms (\ie, \gls{mlp} and \gls{lstm} ) for classification purposes. Moreover, the authors propose to use the underlying TCP variant as a practical feature for improving classification accuracy. The authors in \cite{SONG20191} compare the performance of the ML-based techniques, such as k-nearest neighbors and Decision Tree, with the traditional commercial rule-based strategy for operating systems fingerprinting. The simulation result demonstrates the superiority of the learning-based techniques to the traditional method. Lastovicka \textit{et al.} in \cite{lastovicka2018passive} investigate the performance of the three famous operating system fingerprinting techniques, including user-agent, TCP/IP parameters fingerprint, and specific domains communication. Performance measures reveal that the method based on user-agents provides better performance than its counterparts.}
 \end{itemize}

\section{Overview on \gls{al}} \label{sec:al}
 
A supervised machine learns to discriminate the different traffic classes by being trained on labeled training data. While capturing large quantities of network data is relatively easy, analysing the data by \gls{ml} techniques can be a very time-consuming, expensive, or human-labor intensive process. This is mainly because of the complexity of \gls{ml} techniques or the shortage of labeled data resulting in inefficient training. 
In order to reduce the number of needed labeled examples and, consequently, reduce the effect of ground truth challenge, \gls{al} can be used to facilitate labeling. 

\begin{figure}[b!]
 \centering
 \includegraphics[width=7cm]{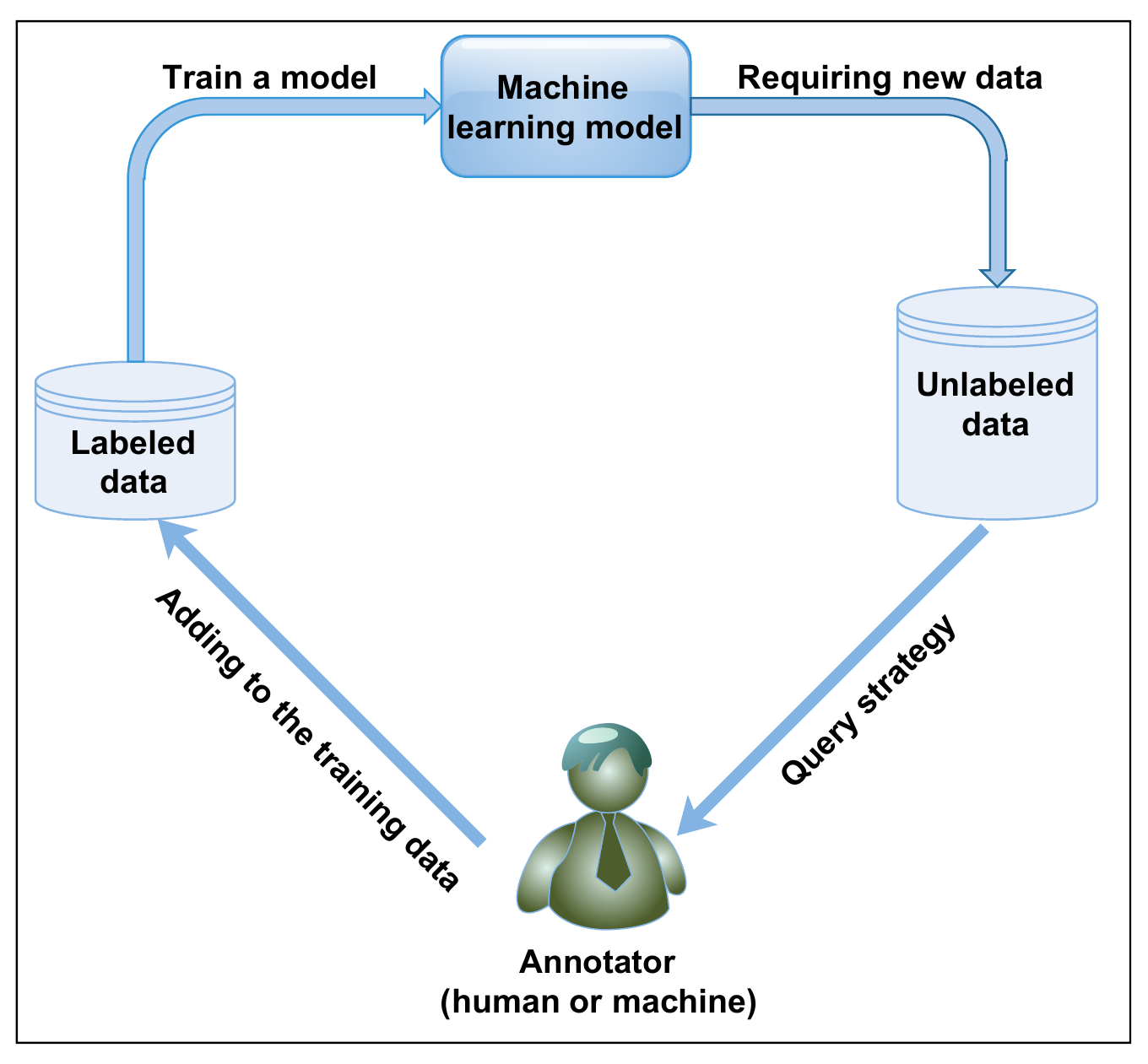}
 \caption{Graphical description of active learning.}
 \label{fig:active learning structure}
\end{figure}

\gls{al} systems can participate in the gathering and selection of training instances, such that only the most informative examples are required to be labeled. 
Using \gls{al}, a learner follows an iterative strategy in which it interacts with an oracle to choose the most useful data instances to be labeled, thereby, it reduces the cost of data labeling by using only a few labeled examples to deliver satisfactory performance in a reasonable time. The \gls{al} paradigm is illustrated in Fig. \ref{fig:active learning structure}, in which the three core components are: \textcolor{black}{\textit{query strategy}, \textit{annotator}, and \textit{\gls{ml} model}}. The query strategy is responsible for choosing unlabeled data according to a pre-defined policy. A label is then provided for the selected data by a human/machine annotator, and the data is added to the set of training instances. Afterwards, the model is updated, and the process repeated as long as new data is available, or a stopping criterion is satisfied. \textcolor{black}{Different stopping criteria can be defined to end this iterative process, such as reaching the desired accuracy, running time, or a maximum number of queries, which can directly affect the performance of using \gls{al}.}


\begin{figure*}[t]
 \centering
 \includegraphics[width=16cm]{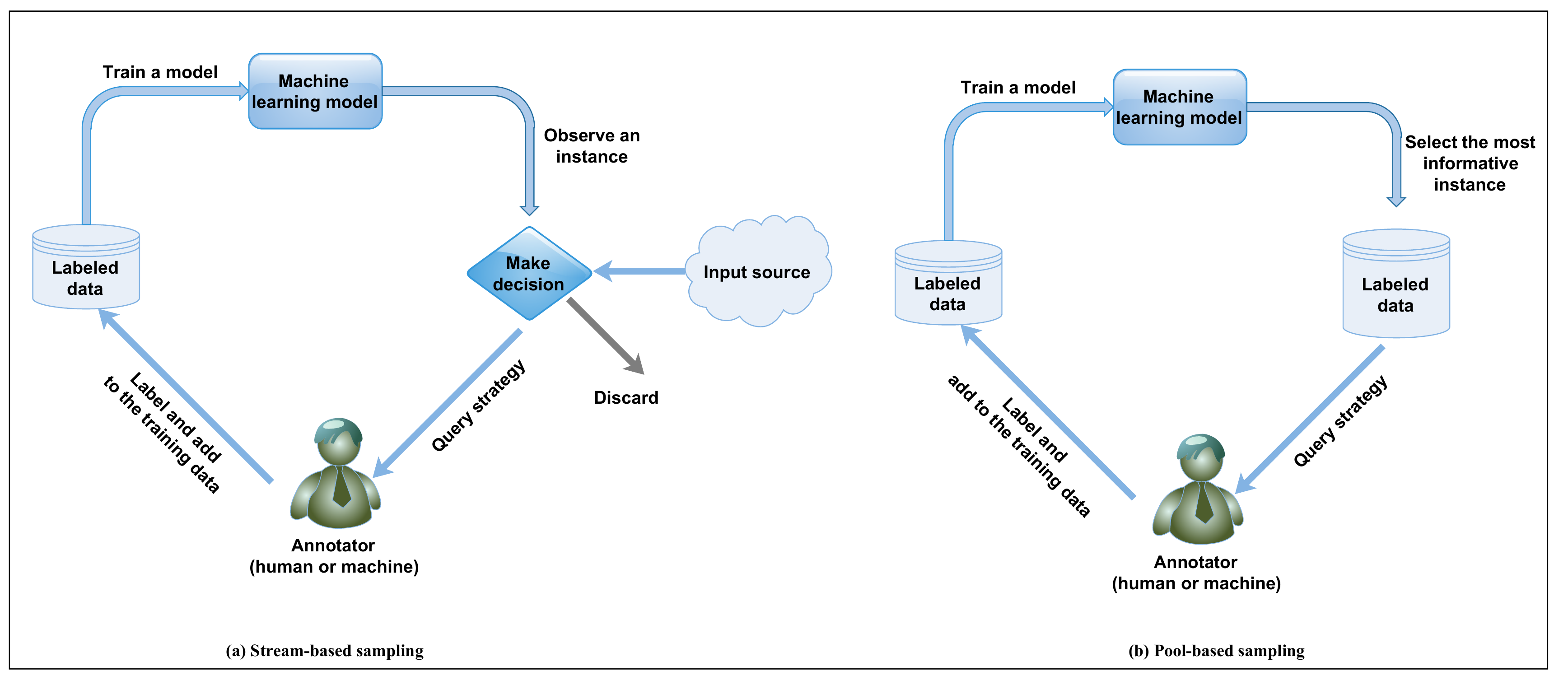}
 \caption{(a) Stream-based selective sampling, and (b) Pool-based sampling.}
 \label{fig:bothAL}
\end{figure*}

There are mainly two \gls{al} scenarios to consider, namely, \textit{stream-based} selective sampling and \textit{pool-based} sampling (presented in Fig. \ref{fig:bothAL}). In the former, the distribution of unlabeled instances is known, and the instances are considered one at a time. The learner then observes each instance in sequence and decides whether the instance should be labeled or discarded.
\textcolor{black}{\gls{al} is a promising technique to alleviate the challenge of streaming-based learning scenarios \cite{5440901}, \cite{6414645}. 
\gls{al} algorithms designed for streaming scenarios can control the labeling process and gradually perform this process over time \cite{9012675}. Using this strategy, it is expected that the labeling process will be in balance and the algorithms will detect the changes.}
 In the case of pool-based sampling, a pool of unlabeled data is provided, and the aim of the learner is to select the most informative instances from the pool to be labeled by the annotator. \textcolor{black}{ Pool-based sampling is attractive for many real-world learning scenarios as it is possible to collect a large body of unlabeled data at once. Pool-based sampling presumes that a limited amount of labeled data and a big pool of unlabeled data are available.}

\subsection{Active learning query strategies}
\label{sec:querystrategies}
The fundamental question in \gls{al} is that what is the most effective strategy for querying data instances? 
In \gls{ntc} applications, different query strategies can be used based on various network circumstances, \eg new unknown flows, changes in the behavior of network traffic, and discovering unclassified network traffics. We first, introduce the most well-known query strategies of \gls{al} widely used in literature and then evaluate the performance of the query strategies in Section \ref{sec:performance}.
\textcolor{black}{Note that in \gls{ml} terminology, hypothesis space refers to the all possible legal hypotheses, where a hypothesis is a particular computational model that best explains the target data in supervised \gls{ml}. In active learning settings, a query strategy can search the hypothesis space through testing unlabeled samples to reduce the number of legal hypotheses under attention.}

\begin{itemize}
 \item \textbf{\gls{unc}:} 
 In \gls{unc}, a learner prefers to label the instances where the model is most uncertain about the class of the example. The idea behind the strategy is that those examples on which the model exhibits the most degree of uncertainty are most likely to improve the performance of the model over time. Different criteria, also called \textit{uncertainty strategies}, for measuring uncertainty, have been proposed including posterior probability, smallest margin, and entropy \cite{settles2008curious}. Entropy is one of the most popular uncertainty strategies in many \gls{al} problems.
 In an \(n\)-class classification problem, assume the estimated probabilities of the \(n\) classes are $p_1,…,p_n$, respectively.
 Given the currently labeled data instances, the entropy is defined as $E(\overline{\rm X})=-\sum_{i=1}^{n}p_i.\log	(p_i)$.
 \textcolor{black}{Given this expression, a larger value of the entropy means a higher level of uncertainty. Accordingly, this objective function can be considered as a maximization problem.}
 
 \item \textbf{\gls{qbc}:} In \gls{qbc}, an \gls{al} system consists of a committee of different learners trained on the current labeled data. These learners are then used to make a prediction on the labels of unlabeled data. The instances for which the committee members disagree the most on the correct label are selected for labeling. Then, the committee of learners will use the new labeled data examples for training purposes. 
 The QBC strategy creates wider diversity than \gls{unc} because it considers the differences in the predictions of several different learners, instead of measuring the level of uncertainty of labeling using only a single learner. However, the technique for measuring the disagreement is often similar for both query strategies \cite{fundament}.
  \textcolor{black}{In the \gls{qbc} strategy, the vote entropy and KL-divergence metrics are usually applied to measure the disagreement. In the literature, to construct a committee of learners, two major approaches have been proposed.
 In the former, one can change the parameters/hyperparameters of a particular model (\eg, by sampling) in order to generate different models and, consequently, the committee models (or learners).
 In contrast, in the latter, the committee is built by a bag of different learners (\ie, ensemble of learners).}
 
 \item \textcolor{black}{\textbf{Learning Active Learning (LAL):} The main idea behind this strategy is to train a regressor that forecasts the \gls{eer} for an instance in a specific learning state.
 \textcolor{black}{Indeed, this technique formulates the query strategy of unlabeled data as a regression problem.}
 Then, regarding a trained classifier and its output for specific unlabelled instance,
 the \gls{lal} forecasts the decrease in generalization error that can be reached by labeling that instance. The interested readers are referred to read \cite{konyushkova2017learning} for details. 
 }
 \item \textcolor{black}{\textbf{Random:} 
 It refers to the conventional supervised learning scheme in which instances are randomly selected to be labeled. Since data labeling is an expensive procedure, random sampling may not lead to the best learner, \textcolor{black}{ especially when the query of each sample is costly, and consequently, few labels will finally be available \cite{fundament}.}
 \item \textbf{Information Density (Density):}
 Uncertainty sampling, \gls{qbc}, and \gls{lal} query strategies }are all prone to choosing outliers or unrepresentative instances and, consequently, this can lead to sub-optimal queries.
 A solution is to use the representativeness of an instance to ensure the selected instances resemble the overall distribution. When considering whether to query an instance, a combination of representativeness and the informativeness instances is typically used \textcolor{black}{ {\cite{settles2008analysis}. In the density query, to measure the representativeness of a data instance, the closeness of the data instance to all other data instances is often considered.}}
\end{itemize}

\section{Active learning for network traffic classification} \label{sec:active NTC}
 As explained in Section \ref{sec:ntc}, \gls{ntc} has attracted much interest in recent years and different ML methods have been proposed to solve the \gls{ntc} problem. However, most of these methods suffer from various challenges such as requiring a large amount of fully labeled data, existence of a considerable amount of semi-labeled or unlabeled data in real-world \textcolor{black}{network scenarios, and complex, costly, and time-consuming methodology for data labeling.}
 \textcolor{black}{Providing labels to data instances is especially challenging for \gls{ntc} techniques, because one must consider several requirements in terms of traffic data granularity in order to satisfy the desired traffic classification objectives. One can, for example, refer to classes on the application level (\ie Skype or Facebook), protocols level (\ie TCP or HTTP), or at the service group level (\ie browsing or streaming) as typical examples of data granularity \cite{velan2015survey}.}
 Moreover, updating a traffic classification method is time-consuming. However, updating the models may be needed to increase the accuracy of the method or recognize new applications, protocols, or protocol versions. The update is essentially performed using new labeled data. 
 
\gls{al} is a promising research field in this context as it greatly reduces the cost of training and dramatically speeds up the learning phase \cite{settles2011theories}. This is advantageous to ML-based traffic classification methods to better satisfy the \textcolor{black}{aforementioned requirements, precisely data requirements and the need for updating to identify new types of traffic through attaching labels on the most informative instances}
and the need for updating to identify new types of traffic. 

\subsection{Advantages of using \gls{al} for \gls{ntc} purposes }
 \gls{al} is potentially a good candidate to perform \gls{ntc}. Below, we summarize the advantage of using \gls{al} techniques in the field of \gls{ntc}:

\begin{itemize}
 \item \textit{Less amount of data needed for labeling}: As mentioned before, most conventional networks generate unlabeled and semi-labeled data. Meanwhile, one of the key challenges to use the learning-based techniques for \gls{ntma} is the lack or limited accessibility of labeled instances. Moreover, data labeling is not often a straightforward procedure and can raise the cost in terms of time, human effort, and the computational overhead. Other than that, if data labeling is performed manually or by online tools, it can reduce the data quality, since not all data instances are informative. 
 \gls{al} can \textcolor{black}{tackle this concern} by labeling only the most informative instances.
 To this end, a comprehensive set of querying strategies in \gls{al} has been proposed to determine the quality of instances for labeling \cite{settles1648active}.
 
 \item \textit{Concept Drift}: Due to high dynamicity of computer networks, \gls{ml} techniques must be re-trained frequently because of various reasons, \eg new network behavior and new classes of network traffic \cite{shahraki2020boosting}. In most \gls{ml} techniques, such as \gls{dl}, retraining a model from scratch is a resource-intensive task in terms of time and power computation in addition to their need for huge amount of new data samples. Most well-known \gls{ml} techniques become useless in \gls{ntc} as the network cannot be unattended for a long time due to retraining purposes. \gls{al} is able to (re-)train the models very fast with high accuracy \textcolor{black}{by continuous provisioning of new labeled instances.}
 This is demonstrated in Section \ref{sec:performance} where AL performance is evaluated with regard to the training time.
 
 \item \textcolor{black}{\textit{Dealing with the shortage of labeled data samples}: In case of retraining, the number of labeled samples to train the model is very limited due to the cost of labelling process, \eg, time, complexity, need for domain knowledge, etc.
 Most \gls{ml} models, \eg \gls{dl}, need a considerable amount of data to train. As shown in Section \ref{sec:performance}, \gls{al} can train the model with a high accuracy using a limited number of data samples.}
 
 \item \textit{Incremental Learning}: Although \gls{al} is not essentially considered as an online learning technique, using the stream-based sampling can possibly turns it into an incremental learning technique to be adaptable with the nature of highly dynamic networks. As most of conventional and emerging networking paradigms are highly dynamic in different aspects, \gls{al} can be used to learn the behavior of network traffic online. In addition, pool-based sampling can help reduce the time complexity of learning from scratch, as the number of training samples becomes limited. Although labeling is a time-consuming task, using different query strategies based on the network traffic circumstances can reduce the time complexity of learning.
 
 \item \textcolor{black}{\textit{Monitoring incoming stream traffic}: Using passive learning methods for \gls{ntc} tasks, such as security and intrusion detection is no longer reasonable, as these methods cannot handle \textcolor{black}{changes in the statistical characteristics of the target data (\ie, concept drift).}}
 \textcolor{black}{To address this issue, one can investigate the great abilities of stream-based \gls{al} \cite{yang2011active}.} Several AL-based strategies have been proposed to detect concept drift and instantly adapt to evolving characteristics of data \cite{krawczyk2018combining} \cite{zhang2018online} \cite{costa2018drift}.
 
 \item Addressing \textit{Theory of network}: In Internet Engineering Task Force 97 (IETF97)\textcolor{black}{\footnote{https://www.ietf.org/blog/reflections-ietf-97/}}, the challenge is introduced as networks suffer from the lack of a unified theory that can be applied to all networks. It means that the behaviors of different networks are various based on their topology, equipment, scale, applications, etc. \textcolor{black}{Theory of Network causes an important problem that \gls{ml} techniques should be trained for each network separately. AL can be considered as a suitable online learning choice in such cases thanks to its ability to be learned by a limited number of data samples. This is beneficial for highly dynamic networks with a huge volume of starting and stopping network traffics. AL also allows frequent retraining which eliminates the necessity of using representative datasets.} 
 \end{itemize}

\subsection{Literature Review on using \gls{al} in \gls{ntc}}
\textcolor{black}{In this section, we review existing work on the application of \gls{al} in \gls{ntc}.}



Torres textit{et al.} \cite{torres2019active} proposed a botnet detection technique based on \gls{al}. The authors provided a novel \gls{al} strategy to label network traffic that contains normal and botnet traffics. The \gls{al} strategy is used to create a random forest model that benefits from the user's previously-labeled instances. The primary objective of the proposed technique is to help the user in the labeling process. Similarly, the work in \cite{chen2020malware} employed \gls{al} for a security purpose, \ie, malware classification. In this work, \glspl{svm} and \gls{albl} have been combined to tackle the lack of labeled instances in malware detection. The simulation results reveal that using \gls{al} can enhance the performance of classification in terms of accuracy and the quality of labeled instances. In addition, the authors claimed that by using different training algorithms, \eg \glspl{gan}, one can solve issues such as the diversity of security-related datasets. 

The work in \cite{nissim2016aldocx} is another attempt to develop an accurate malware detection system. The system is based on \gls{al}, where a new \gls{sfem} is introduced to extract from docx files. The proposed system is able to identify new unknown malicious docx files. To have an updatable detection model and identify new malicious files, the system benefits from \gls{al} to update and complement the signature database with new unknown malware.

Common cybersecurity attack vectors, such as viruses, botnets, and malware are known for \glspl{ids}. Nevertheless, malicious users continuously create new attacks that can bypass the \glspl{ids}.
Analyzing anomalous behaviors calls for a considerable amount of time and effort. Preparing a significant of labeled data for the training process is both increasingly costly and inefficient, because of the continuous design of new attacks. In this case, one can use \gls{al} to reduce the number of the required labeled instances, while increasing the accuracy of anomaly detection. In \cite{8058397}, a semi-supervised IDS has been designed that works effectively with a small number of labeled instances. The proposed learning algorithm for the IDS benefits from two ML techniques, including \gls{asvm} and Fuzzy C-Means clustering. Furthermore, \cite{8058397} reported that the proposed learning algorithm enables the \gls{ids} to add new training instances with minimum computational overhead.
Due to the fact that domain knowledge is required for the annotations of unlabeled instances, adopting new cost-effective labeling techniques is desired. To this end, the work in \cite{springer111} by Beaugnon \textit{et al.} developed an interactive labeling strategy, namely ILAB, to assist the experts in the labeling process of large intrusion detection datasets. ILAB adopts divide and conquer approach to lower the computation cost. Deka \textit{et al.} \cite{DEKA2019203} investigated the important role of \gls{al} in the selection of more informative instances. Then, they used these instances to train a binary \gls{ids} for \gls{ddos} attack classification. In addition, since there are massive amounts of traffic in modern networks, a parallel computation method has been employed. The authors referred to this fact that using \gls{al} is an efficient technique to keep the cost of the labeling down and avoid keeping redundant training instances. 

Domain-specific anomaly detection has been targeted in \cite{8707963} \textcolor{black}{by establishing} an \gls{al}-based framework of tripartite \gls{al} to interactively detect anomalies. In this framework, a two-stage algorithm is used for labeling instances. In the first stage, the algorithm selects instances based on the uncertainty criterion for labeling. Then, it uses a technique to evaluate and train multiple annotators with the most appropriate instances. Unlike other related works on \gls{al}, this paper investigates human-in-the-loop active \gls{ml}. \textcolor{black}{Another work that adopted AL has been conducted by Shu \textit{et al.} \cite{shu2020generative}. They focus on the investigation of adversarial attacks on ML-based \gls{ids}. More specifically, the authors propose using \gls{al} and generative adversarial networks to evaluate the related threats in \gls{ids} that use learning-based techniques. They highlight that current adversarial attack techniques need a massive amount of labeled data for training purposes. They propose using \gls{al} as a solution to tackle this problem.}

Wassermann \textit{et al.} in \cite{wassermann2019adam} examine two central issues in stream-based ML and online network monitoring, \ie, 1) how to learn in dynamic environments in the presence of concept drifts, and 2) how to learn with a small number of labeled data and, regularly improve a supervised model through new samples. To deal with these issues, the authors propose two stream-based ML algorithms, namely ADAM and \gls{ral}. The \gls{ral} algorithm is based on \gls{al} to reduce the need for ground truth samples for stream-based learning. \textcolor{black}{\textcolor{black}{Then, the authors use the proposed algorithms in continuous network monitoring for attack detection.}}

 As explained above most of existing works focus on the use of \gls{ml} in \gls{ntc} for network security improvement, but there are a few works that use \gls{al}-based \gls{ntc} techniques for other needs. 
 The work in \cite{liu2015active} uses \gls{al} for P2P traffic classification. They proposed P2PTIAL, a new method for \gls{p2p} traffic classification based on \gls{al}, \textcolor{black}{consisting of two parts, an \gls{svm} classifier and uncertainty query strategy.}
 Moreover, to further improve the performance of classification, they adopted filtering and a balanced policy to their method. The main idea behind this filtering is to prevent from unlabeled less informative samples and consequently saving cost in terms of computation and storage space.
 \textcolor{black}{\textcolor{black}{The authors in \cite{yin2019incorporate} use the fusion of \gls{al} and semi-supervised learning for industrial fault detection. They propose using \gls{al} to improve the performance of the fisher discriminant analysis method since it has difficulty when the labeled instances are not satisfactory. The simulation results reveal that \gls{al} and semi-supervised methods can complement each other. Similar work has been performed in \cite{yin2018active}, in which the authors propose an \gls{al} method based on \gls{svdd} for novelty detection in industrial data. \textcolor{black}{ The work in \cite{dong2021multi} introduces an active version of multi-class \gls{svm} algorithm, called cost-sensitive \gls{svm} (CMSVM) to deal with the imbalance problem in network traffic. To this end, CMSVM uses a multi-class SVM technique with \gls{al} which dynamically allocates a weight to class labels. The author claims that the proposed method can decline computation load, increase classification accuracy, and alleviate the imbalanced data problem.
 }}}
 
 A summary of the reviewed papers in this section is provided in Table \ref{tab:NtcAlComparison}. \textcolor{black}{As shown in the table, the use of \gls{al} techniques in \gls{ntc} is basically for the improvement of network security.}

\begin{table*}[]
\centering
\tiny
\caption{\textcolor{black}{Summary of papers on active learning for \gls{ntc}.}}
\label{tab:NtcAlComparison}
\begin{tabular}{|l|l|l|l|p{3cm}|}
\hline
\textbf{Ref.} & \begin{tabular}[c]{@{}l@{}}\textbf{NTC challenge}\end{tabular} & \textbf{Method} & \textbf{Technical contribution}
 \\ \hline
\cite{torres2019active}&\begin{tabular}[c]{@{}l@{}}Botnet detection\end{tabular}&\begin{tabular}[c]{@{}l@{}}Active learning+random forest\end{tabular} &\begin{tabular}[c]{@{}l@{}}Proposes a new \gls{al} strategy to facilitate the network traffic labeling process, especially for security purposes.\end{tabular} \\ \hline
 \cite{chen2020malware}&\begin{tabular}[c]{@{}l@{}}Malware classification\end{tabular}&\begin{tabular}[c]{@{}l@{}}\gls{albl}+ SVMs\end{tabular}&\begin{tabular}[c]{@{}l@{}}Combines \gls{al} by learning and SVM to construct a classifier\\ for malware classification with a small number of labeled samples. \end{tabular} \\ \hline
 \cite{nissim2016aldocx}&Malware detection &\begin{tabular}[c]{@{}l@{}}\gls{al} +\gls{sfem}\end{tabular}&\begin{tabular}[c]{@{}l@{}}Establishes a novel \gls{al}-based framework, namely ALDOCX, to identify new unknown malicious docx files. \end{tabular} \\ \hline
 \cite{8058397}&Intrusion detection&\begin{tabular}[c]{@{}l@{}}\gls{al}+ASVM+Fuzzy C-Means\end{tabular}&\begin{tabular}[c]{@{}l@{}}Provides a semi-supervised method for intrusion detection based on an active version of SVM and fuzzy c-means. \end{tabular} \\ \hline
 
 \cite{springer111}&\begin{tabular}[c]{@{}l@{}}labeling IDS dataset \end{tabular}&\begin{tabular}[c]{@{}l@{}}\gls{al}+ \\divide and conquer \end{tabular} &\begin{tabular}[c]{@{}l@{}}The paper proposes a labelling technique, called ILAB, for network security purposes. The technique works in an interactive fashion\\ to decrease the cost of labeling in terms of workload. \end{tabular}\\ \hline

\cite{DEKA2019203}&\begin{tabular}[c]{@{}l@{}}\gls{ddos} attack classification \end{tabular}&\begin{tabular}[c]{@{}l@{}}\gls{al} +SVM\end{tabular}& \begin{tabular}[c]{@{}l@{}}Provides a parallel cumulative ranker method to rate the features of a network traffic dataset. Then, leverages \gls{al} to select\\ the most informative samples for training an SVM classifier. \end{tabular}\\ \hline

\cite{8707963} &Anomaly detection& \begin{tabular}[c]{@{}l@{}}Tripartite \gls{al} 
\end{tabular} &\begin{tabular}[c]{@{}l@{}}Establishes a tripartite \gls{al} framework to detect anomalies in an interactive manner through crowd-sourced labels. \end{tabular}\\ \hline

 \cite{wassermann2019adam}&\begin{tabular}[c]{@{}l@{}}Network attacks\\detection \end{tabular}&\begin{tabular}[c]{@{}l@{}}Reinforcement learning\\+\gls{al} \end{tabular}&\begin{tabular}[c]{@{}l@{}}The paper proposes two ML algorithms for the stream-based environments to tackle some common problems in these environments,\\ such as the lack of labeled samples, concept drifts, and having an updated model. \end{tabular}\\ \hline

 \cite{liu2015active}&\begin{tabular}[c]{@{}l@{}}P2P traffic\\classification \end{tabular}&\begin{tabular}[c]{@{}l@{}}\gls{al}+SVM\end{tabular} &\begin{tabular}[c]{@{}l@{}}One of the first works that used \gls{al} for P2P traffic classification to tackle the data labeling challenge.\end{tabular} \\ \hline
 
\cite{lastovicka2018passive}&\begin{tabular}[c]{@{}l@{}}Fault detection \end{tabular}&\begin{tabular}[c]{@{}l@{}}\gls{al}+semi-supervised\end{tabular} &\begin{tabular}[c]{@{}l@{}}One of the first works that combine \gls{al} with semi-supervised learning to improve the performance of the fisher discriminant analysis method.\end{tabular}
\\ \hline
 
 \cite{yin2018active}&\begin{tabular}[c]{@{}l@{}}Novelty detection \end{tabular}&\begin{tabular}[c]{@{}l@{}}\gls{al}+\gls{svdd}\end{tabular} &\begin{tabular}[c]{@{}l@{}}Proposes to use \gls{al} to solve the problem in \gls{svdd}, i.e., the bad performance when data is massive, and the quality is poor.\end{tabular} \\ \hline
 
 \cite{dong2021multi}&\begin{tabular}[c]{@{}l@{}}Traffic classification \end{tabular}&\begin{tabular}[c]{@{}l@{}}\gls{al}+multi-class \gls{svm}\end{tabular} &\begin{tabular}[c]{@{}l@{}} One of the latest work that adopts \gls{al} and a variant of \gls{svm} for traffic classification.\end{tabular} \\ \hline
 
 \cite{shu2020generative}&\begin{tabular}[c]{@{}l@{}}Adversarial attacks \\evaluation \end{tabular}&\begin{tabular}[c]{@{}l@{}}\gls{al}+generative adversarial networks\end{tabular} &\begin{tabular}[c]{@{}l@{}} One of the first works that use \gls{al} to deal with the lack of labeled data in the investigation of adversarial threats against ML-based \gls{ids}.\end{tabular} \\ \hline
\end{tabular}
\end{table*}
\section{Empirical evaluation of \gls{al} in \gls{ntc}}
\label{sec:performance}
\textcolor{black}{In this section, as a technical survey, we provide useful insights into the performance of \gls{al} and evaluate the most effective and widespread query strategies for \gls{al} settings as the most important \textcolor{black}{aspect of \gls{al} in \gls{ntc}.}} 

As mentioned in Section~\ref{sec:active NTC}, the two important challenges in using \gls{ml} for \gls{ntc} are the need for retraining, and shortage of data samples. 
Our evaluation work is conducted based on these challenges. In particular, we evaluate the performance of \gls{al}-based classification with respect to the training time and shortage of data samples. 
To this end, two examples of using the active form of learning for \gls{ntc} are studied.
It should be noted that we do not consider any labeling technique in the performance evaluation. We assume that all existing samples are labeled and hence, labelling is out of scope of our performance evaluation. However, in real-world applications, the labeling time should be considered. 

\begin{figure}
 \centering
 \begin{subfigure}[b]{0.45\textwidth}
 \centering
 \includegraphics[width=\textwidth]{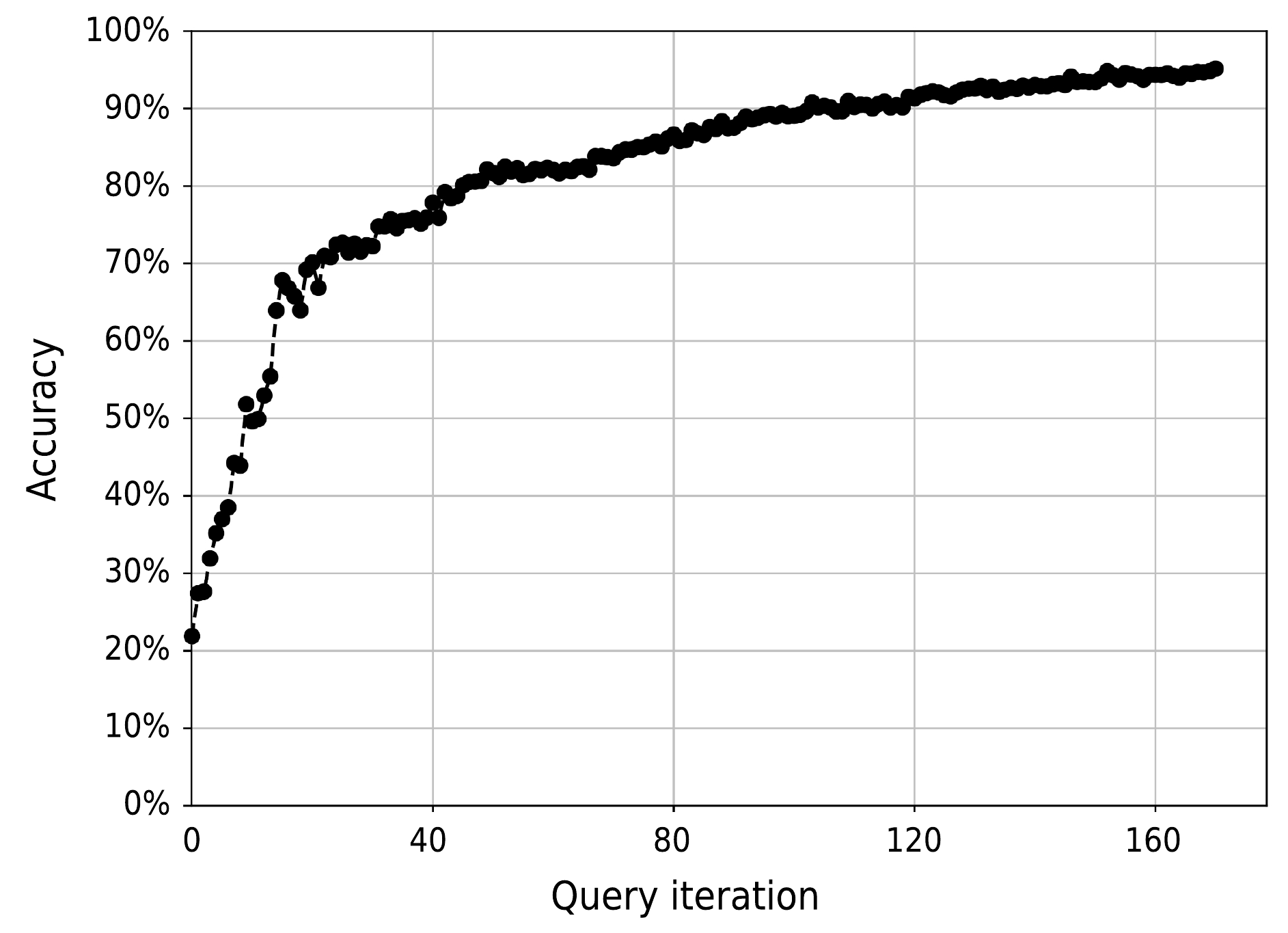}
 \caption{Stream-based scenario}
 \label{fig:my_label44}
 \end{subfigure}
 \hfill
 \begin{subfigure}[b]{0.45\textwidth}
 \centering
 \includegraphics[width=8.2cm,height=6cm]{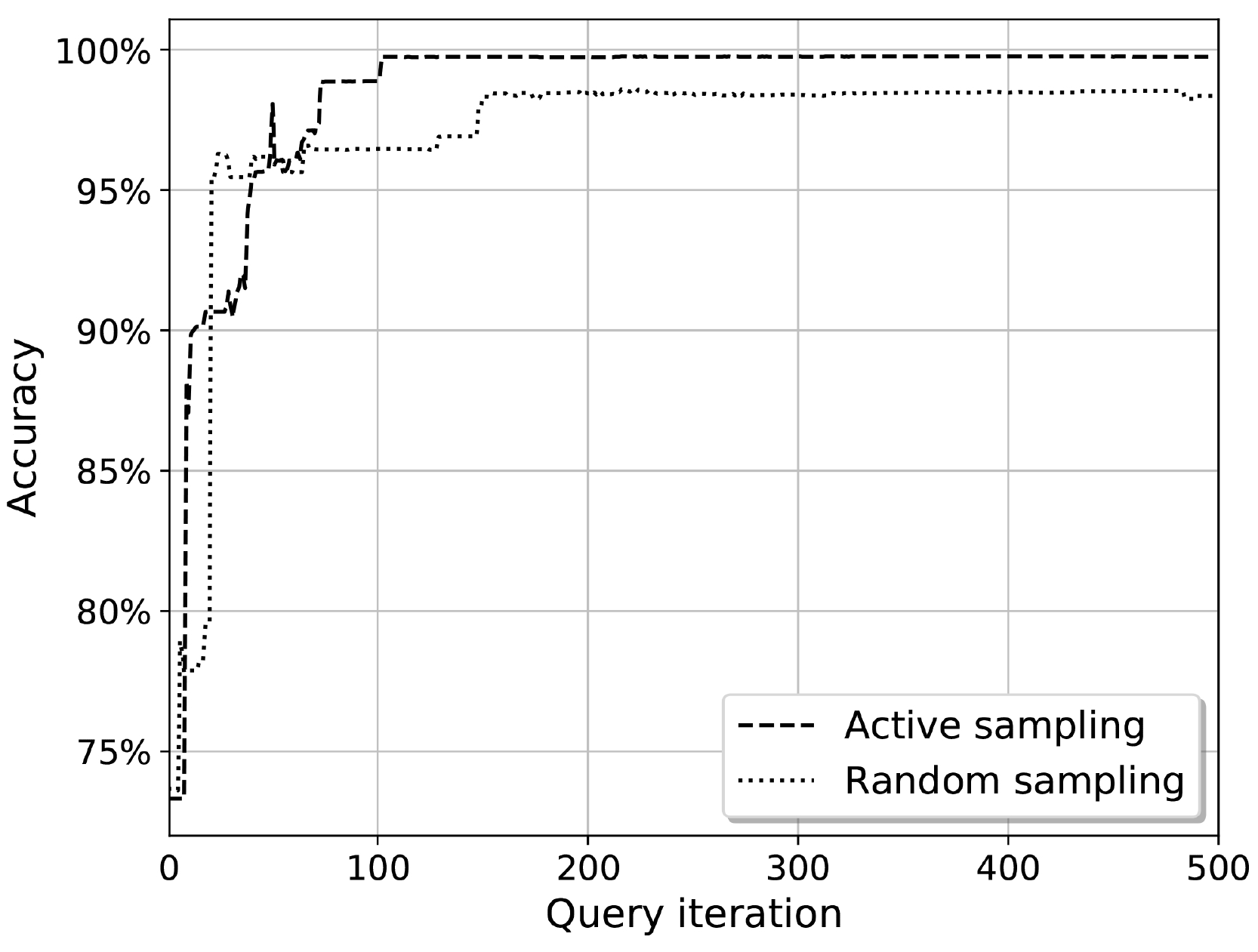}
 \caption{Pool-based scenario}
 \label{fig:my_label55}
 \end{subfigure}
 \caption{\textcolor{black}{Percentage of the classification accuracy for stream-based and pool-based scenario on the Cambridge dataset.}}
\end{figure}

In the first example, we implement a stream-based \gls{al} model for a \gls{ntc} task as shown in Fig. \ref{fig:my_label44}. To this end, we use the Cambridge network traffic dataset as a benchmark dataset and Random Forest as the classifier \cite{moore2005internet}. The dataset has been captured by the Computer Laboratory of the University of Cambridge at various time points of the days from several hosts on three institutions with about 1000 users. This dataset consists of 12 common categories of traffic flow, including WWW, MAIL, ATTACK, P2P, SERVICES, DATABASE, INTERACTIVE, MULTIMEDIA, GAMES, FTP-CONTROL, FTP-DATA, and FTP-PASV. In the stream-based \gls{al}, the unlabeled instances are presented one-by-one to the model. Then, for each observed instance, the learner has to query from the annotator for its label if it recognizes the given instance as a useful instance for the model. For example, one can recognize an instance as a useful instance if the model's prediction is uncertain, since asking for its label may resolve this uncertainty. Different strategies have been provided for stream-based \gls{al} \cite{cheng2013feedback}. Most of these strategies select informative instances according to a single criterion. Considering a single criterion for instance selection may be problematic for \gls{al}, especially for the stream-based \gls{al}, in which a subset of instances selected to be labeled cannot properly show the original distribution of the dataset. 

As shown in Fig. \ref{fig:my_label44}, we measure the performance of the stream-based \gls{al} for \gls{ntc}. It can be seen that the model provides around 30\% accuracy on the initial training dataset. Next, the model chooses useful incoming instances and uses these instances as the new training data. The learning process can be continued until the model reaches the desired accuracy threshold. In our implementation, we set 95\% as the threshold. 
As another example, we consider a pool-based sampling scenario (non-stream) for \gls{ntc}. Under this scenario, a small set of labeled traffic data $\Gamma$ and a large set of unlabeled traffic data $U$ \textcolor{black}{are chosen} so that $ |\Gamma|$ $ \ll$ $ |U| $. As can be seen in Fig. \ref{fig:my_label55}, \gls{al} gives better performance than the random sampling method (\ie passive learner). In active sampling, instances are selectively chosen from the pool of instances. As mentioned, different query strategies have been provided to choose the most informative instances. By doing so, one can \textcolor{black}{reduce} the need for labeled instances and achieve higher accuracy than a passive model. 

 
\begin{figure*}
 \centering
 \begin{subfigure}[b]{0.32\textwidth}
 \centering
 \includegraphics[width=\textwidth]{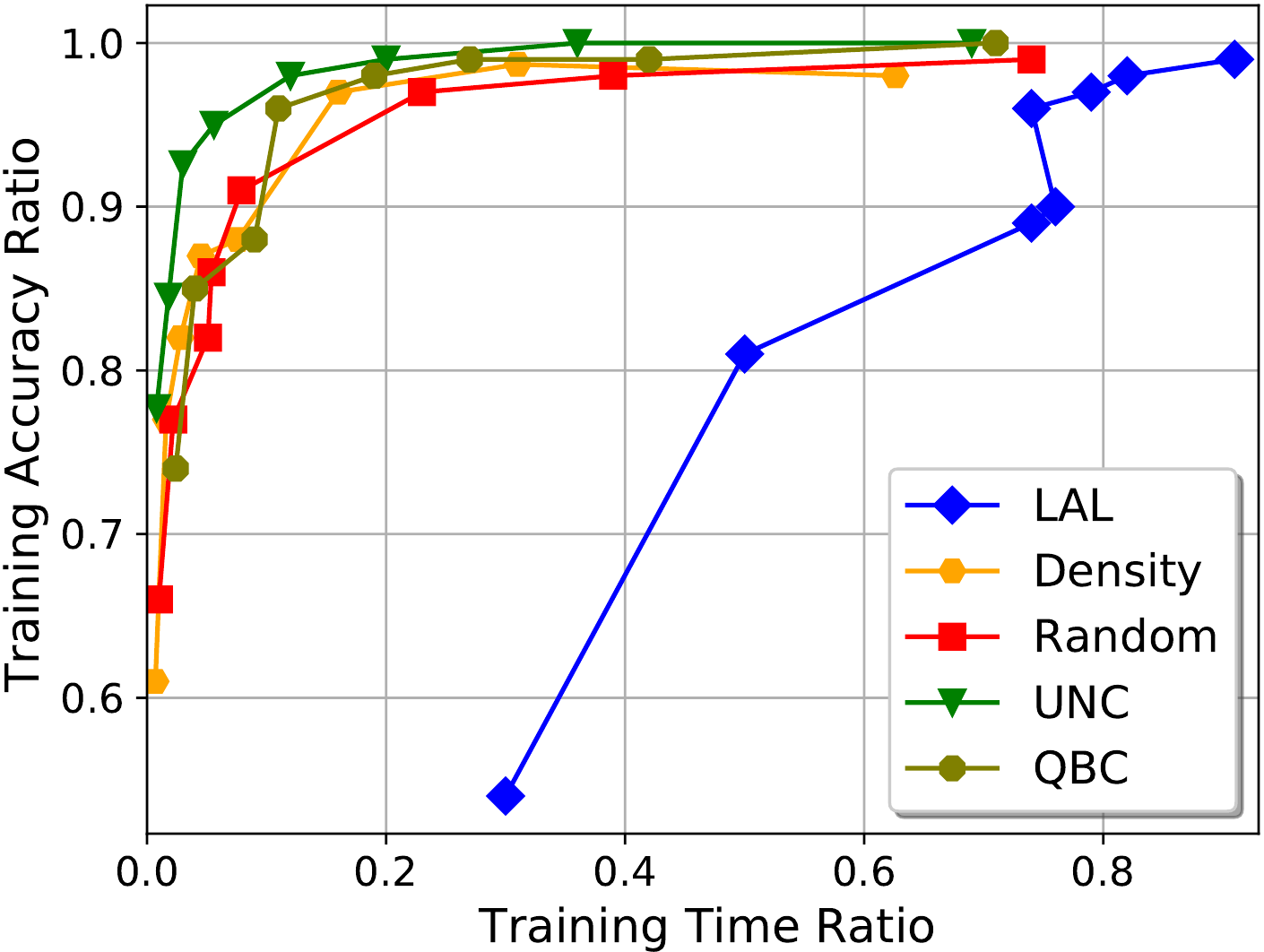}
 \caption{TRAbID dataset}
 \label{fig:trabid}
 \end{subfigure}
 \hfill
 \begin{subfigure}[b]{0.32\textwidth}
 \centering
 \includegraphics[width=\textwidth]{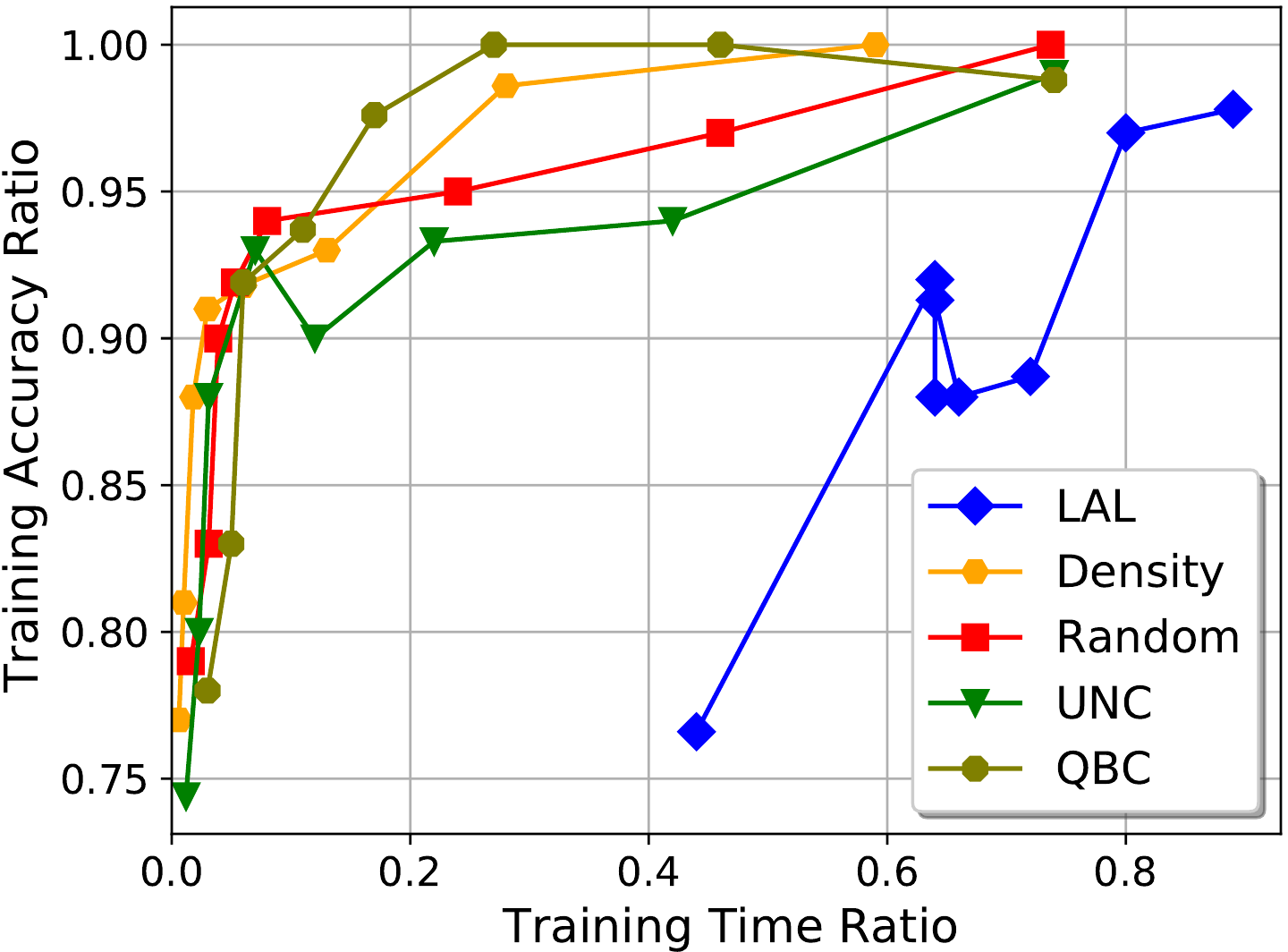}
 \caption{VPN-nonVPN dataset}
 \label{fig:vnv}
 \end{subfigure}
 \hfill
 \begin{subfigure}[b]{0.32\textwidth}
 \centering
 \includegraphics[width=\textwidth]{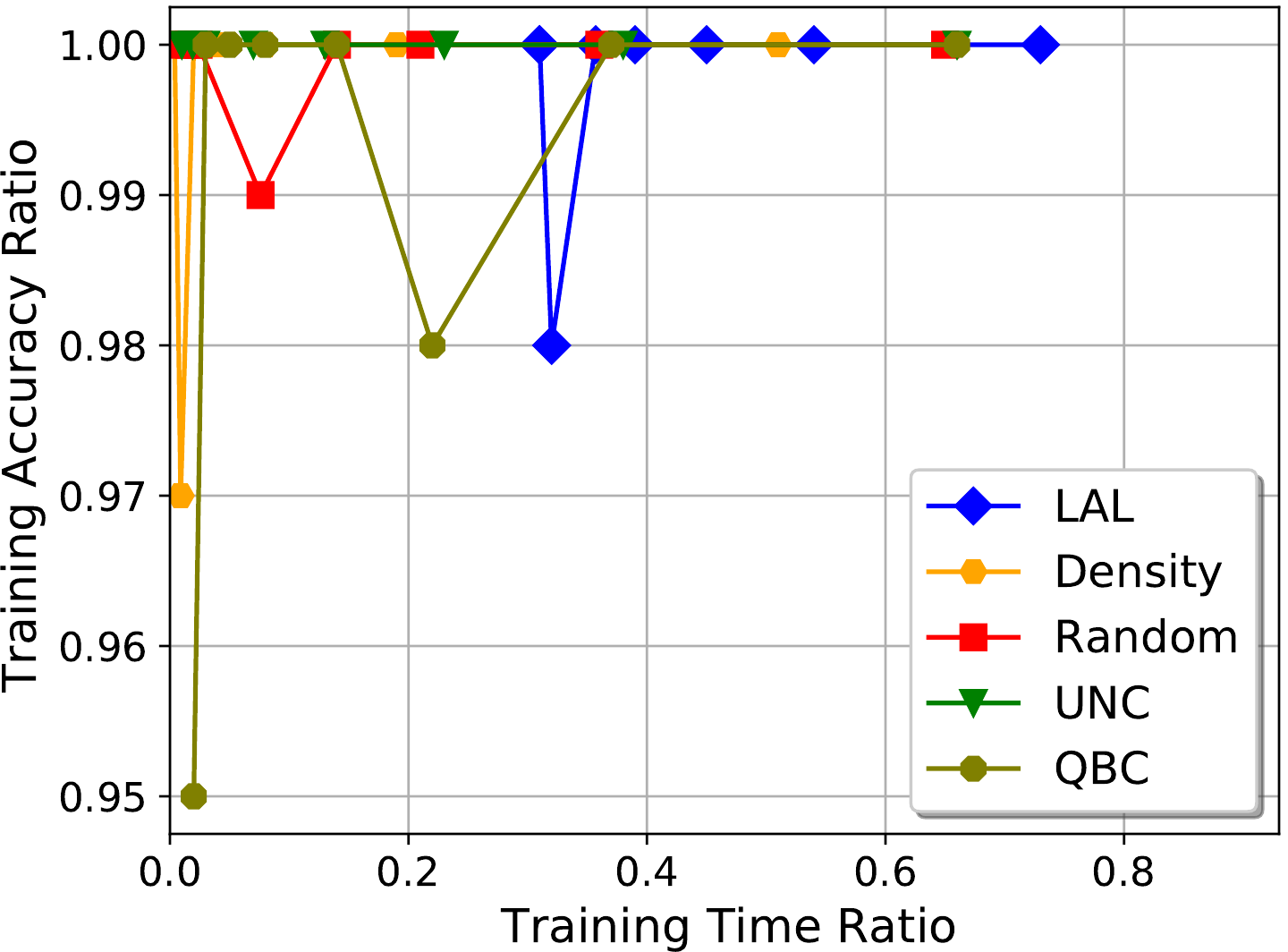}
 \caption{Tor-nonTor dataset}
 \label{fig:tnt}
 \end{subfigure}
 
 \caption{\textcolor{black}{Experimental results on \gls{al} by using different query strategies on NTC datasets. Each point in the graphs shows what percentage of the dataset has been used for training, ranging from \%0.5, \%1, \%2, \%4, to \%64, as indicated in Tables \ref{tbl:trabid)}, \ref{tbl:vpn} and \ref{tbl:tornotor}.}} 
 \label{fig:three graphs}
\end{figure*}

Fig. \ref{fig:three graphs} shows the evaluation results of the training accuracy ratio and training time ratio of the aforementioned query strategies based on three well-known benchmark datasets, \textcolor{black}{\ie, TRAbID \cite{viegas2017toward}, VPN-nonVPN \cite{draper2016characterization}, and Tor-nonTor \cite{lashkari2017characterization}}. As there are more than three benchmark datasets in this field, we selected the datasets based on their volume of data to provide small, medium, and big datasets. In all datasets, we first shuffle the data and then select 0.5\%, 1\%, 2\%, 4\%, 8\%, 16\%, 32\%, 64\% of the whole dataset as a subset to train and test the model by different query strategies. Moreover, to clarify the figures, we elaborate them in Tables \ref{tbl:trabid)}, \ref{tbl:vpn} and \ref{tbl:tornotor}. The X-axis of the figures shows the training time for the full dataset. As an example to interpret tables and figures, when 0.5\% of the dataset is injected to train the model by using the QBC query strategy, it takes about 3.4 seconds to achieve 74\% of accuracy. In the case of the tables, we calculate the Time and Accuracy for each subset by using different query strategies in various benchmark datasets. In addition, to calculate the Training Accuracy Ratio (TAR) and the Training Time Ratio (TTR) we use Eq. \ref{eq:tar} and Eq. \ref{eq:ttr} as below:
\begin{equation}
\scriptsize
 \label{eq:tar}
 TAR=\frac{\text{Accuracy of the model trained on the subset}}{\text{Accuracy of the model trained on the full dataset}}
\end{equation}
\begin{equation}
\scriptsize
 \label{eq:ttr}
 TTR=\frac{\text{Subset training time+ subset selection time}}{\text{Full dataset training time}}
\end{equation}

Fig. \ref{fig:trabid} and Table \ref{tbl:trabid)} show the results of TRAbID (smallest evaluated dataset). The original number of samples of the dataset is 9159 and its original features are 43, but based on domain knowledge solution, we selected 41 features to train and test.
As shown in the table and figure, Random is the fastest but for a small number of samples, it can not achieve high accuracy. UNC achieves considerable accuracy in a reasonable time compared to other query strategies. LAL is the slowest query strategy as it can not achieve high accuracy by a small number of samples.

Fig. \ref{fig:vnv} and Table \ref{tbl:vpn} show the results of the VPN-nonVPN that has 18759 features. Based on the domain knowledge, we select 22 \textcolor{black}{features} out of 24. Same as TRAbID, Random, and UNC are the fastest query strategist that can achieve considerable accuracy. The most important difference between VPN-nonVPN and TRAbID results is the training time, not the accuracy. It means that by having the same amount of data to train, the evaluated query strategies can achieve the same accuracy, but training times are very different. Fig. \ref{fig:tnt} and Table \ref{tbl:tornotor} show the results of the evaluation on the Tor-nonTor dataset. The whole dataset contains 84194 \textcolor{black}{samples and the original} number of features is 28, but based on the domain knowledge, 18 of them are selected to test and train. Same as for the other datasets, Random and then UNC are the fastest query strategies to achieve high accuracy. LAL is very slow, but compared to other query strategies, it can achieve the same accuracy by having the same percentage of the whole dataset.

As a final result, Random is the fastest query strategy but achieves low accuracy. LAL is the slowest query strategy, but it has a good performance in case of accuracy. Comparing the time and accuracy, UNC is the most adequate query strategy to achieve appropriate accuracy in a reasonable time. In the case of the performance of \gls{al}, the results show that by using the appropriate query strategy, \gls{al} can achieve high accuracy in a limited training time. The results show that \gls{al} is appropriate to be used in \gls{ntc} models as it can be re-trained very fast with a limited number of samples.
\begin{table*}[]
\tiny	
\caption{The results on TRAbID dataset. }
\label{tbl:trabid)}
\begin{tabular}{|c|c||c|c||c|c||c|c||c|c||c|c||c|c||c|c||c|c||c|c||}
\hline
\rowcolor[HTML]{656565} 
{\color[HTML]{FFFFFF} \begin{tabular}[c]{@{}c@{}}Query\\  Strategy\end{tabular}} & {\color[HTML]{FFFFFF} \begin{tabular}[c]{@{}c@{}}Evaluated\\  Parameters\end{tabular}} & \multicolumn{2}{c||}{\cellcolor[HTML]{656565}{\color[HTML]{FFFFFF} 0.5\%}}                              & \multicolumn{2}{c||}{\cellcolor[HTML]{656565}{\color[HTML]{FFFFFF} 1\%}}                                 & \multicolumn{2}{c||}{\cellcolor[HTML]{656565}{\color[HTML]{FFFFFF} 2\%}}                               & \multicolumn{2}{c||}{\cellcolor[HTML]{656565}{\color[HTML]{FFFFFF} 4\%}}                                & \multicolumn{2}{c||}{\cellcolor[HTML]{656565}{\color[HTML]{FFFFFF} 8\%}}                                & \multicolumn{2}{c||}{\cellcolor[HTML]{656565}{\color[HTML]{FFFFFF} 16\%}}                              & \multicolumn{2}{c||}{\cellcolor[HTML]{656565}{\color[HTML]{FFFFFF} 32\%}}                               & \multicolumn{2}{c||}{\cellcolor[HTML]{656565}{\color[HTML]{FFFFFF} 64\%}}                               & \multicolumn{2}{c||}{\cellcolor[HTML]{656565}{\color[HTML]{FFFFFF} Full}}                        \\ \hline
\cellcolor[HTML]{656565}{\color[HTML]{FFFFFF} }                                  & Time$|$ACC                                                                               & 3.4                                                & 0.74                                              & 6.9                                                & 0.85                                               & 12.6                                              & 0.88                                              & 16.4                                               & 0.96                                             & 26.9                                               & 0.97                                             & 38                                                & 0.98                                             & 59.4                                              & 0.98                                              & 100.4                                              & 0.99                                             & 139.8                                          & 0.99                                          \\ \cline{2-20} 
\rowcolor[HTML]{C0C0C0} 
{\cellcolor[HTML]{656565}{\color[HTML]{FFFFFF} QBC}}             & \begin{tabular}[c]{@{}c@{}}TAR$|$TTR\end{tabular}                               & 0.024                                              & 0.74                                              & 0.04                                               & 0.85                                               & 0.09                                              & 0.88                                              & 0.11                                               & 0.96                                              & 0.19                                               & 0.98                                              & 0.27                                              & 0.99                                              & 0.42                                              & 0.99                                               & 0.71                                               & 1                                                 & -                                              & -                                              \\ \hline \hline
\cellcolor[HTML]{656565}{\color[HTML]{FFFFFF} }                                  & Time$|$ACC                                                                               & 0.89                                               & 0.77                                              & 2.03                                               & 0.84                                               & 3.8                                               & 0.92                                              & 6.08                                               & 0.95                                             & 13.1                                               & 0.97                                             & 21.8                                              & 0.98                                             & 40.1                                              & 0.99                                               & 75.2                                               & 0.99                                             & 108.8                                          & 0.99                                          \\ \cline{2-20} 
\rowcolor[HTML]{C0C0C0} 
{\cellcolor[HTML]{656565}{\color[HTML]{FFFFFF} UNC}}             & \begin{tabular}[c]{@{}c@{}}TAR$|$TTR\end{tabular}                               & 0.008                                              & 0.77                                              & 0.018                                              & 0.84                                              & 0.03                                              & 0.92                                             & 0.056                                              & 0.95                                              & 0.12                                               & 0.98                                              & 0.2                                               & 0.99                                              & 0.36                                              & 1                                                  & 0.69                                               & 1                                                 & -                                              & -                                              \\ \hline \hline
\cellcolor[HTML]{656565}{\color[HTML]{FFFFFF} }                                  & Time$|$ACC                                                                               & 0.73                                               & 0.65                                             & 1.6                                                & 0.76                                               & 3.7                                               & 0.81                                              & 3.9                                                & 0.85                                              & 5.7                                                & 0.9                                               & 16.8                                              & 0.96                                             & 28.5                                              & 0.96                                              & 53.5                                               & 0.975                                             & 71.7                                           & 0.98                                          \\ \cline{2-20} 
\rowcolor[HTML]{C0C0C0} 
{\cellcolor[HTML]{656565}{\color[HTML]{FFFFFF} Random}}         & \begin{tabular}[c]{@{}c@{}}TAR$|$TTR\end{tabular}                               & 0.01                                               & 0.66                                              & 0.02                                              & 0.77                                               & 0.05                                             & 0.82                                              & 0.054                                              & 0.86                                              & 0.07                                              & 0.91                                              & 0.23                                              & 0.97                                              & 0.39                                              & 0.98                                               & 0.74                                               & 0.99                                              & -                                              & -                                              \\ \hline \hline
\cellcolor[HTML]{656565}{\color[HTML]{FFFFFF} }                                  & Time$|$ACC                                                                               & \multicolumn{1}{l|}{1.7}                           & \multicolumn{1}{l||}{0.6}                          & \multicolumn{1}{l|}{4}                             & \multicolumn{1}{l||}{0.76}                          & \multicolumn{1}{l|}{6.8}                          & \multicolumn{1}{l||}{0.81}                        & \multicolumn{1}{l|}{10.9}                          & \multicolumn{1}{l||}{0.85}                        & \multicolumn{1}{l|}{18.4}                          & \multicolumn{1}{l||}{0.87}                         & \multicolumn{1}{l|}{39.41}                        & \multicolumn{1}{l||}{0.96}                         & \multicolumn{1}{l|}{76.2}                         & \multicolumn{1}{l||}{0.97}                          & \multicolumn{1}{l|}{151.53}                        & \multicolumn{1}{l||}{0.97}                         & \multicolumn{1}{l|}{242.7}                     & \multicolumn{1}{l||}{0.98}                     \\ \cline{2-20} 
\rowcolor[HTML]{C0C0C0} 
{\cellcolor[HTML]{656565}{\color[HTML]{FFFFFF} Density}}         & \begin{tabular}[c]{@{}c@{}}TAR$|$TTR\end{tabular}                               & \multicolumn{1}{l|}{\cellcolor[HTML]{C0C0C0}0.007} & \multicolumn{1}{l||}{\cellcolor[HTML]{C0C0C0}0.61} & \multicolumn{1}{l|}{\cellcolor[HTML]{C0C0C0}0.016} & \multicolumn{1}{l||}{\cellcolor[HTML]{C0C0C0}0.77} & \multicolumn{1}{l|}{\cellcolor[HTML]{C0C0C0}0.028} & \multicolumn{1}{l||}{\cellcolor[HTML]{C0C0C0}0.82} & \multicolumn{1}{l|}{\cellcolor[HTML]{C0C0C0}0.045} & \multicolumn{1}{l||}{\cellcolor[HTML]{C0C0C0}0.87} & \multicolumn{1}{l|}{\cellcolor[HTML]{C0C0C0}0.076} & \multicolumn{1}{l||}{\cellcolor[HTML]{C0C0C0}0.88} & \multicolumn{1}{l|}{\cellcolor[HTML]{C0C0C0}0.16} & \multicolumn{1}{l||}{\cellcolor[HTML]{C0C0C0}0.97} & \multicolumn{1}{l|}{\cellcolor[HTML]{C0C0C0}0.31} & \multicolumn{1}{l||}{\cellcolor[HTML]{C0C0C0}0.98} & \multicolumn{1}{l|}{\cellcolor[HTML]{C0C0C0}0.62} & \multicolumn{1}{l||}{\cellcolor[HTML]{C0C0C0}0.98} & \multicolumn{1}{l|}{\cellcolor[HTML]{C0C0C0}-} & \multicolumn{1}{l||}{\cellcolor[HTML]{C0C0C0}-} \\ \hline \hline
\cellcolor[HTML]{656565}{\color[HTML]{FFFFFF} }                                  & Time$|$ACC                                                                               & \multicolumn{1}{l|}{91.24}                         & \multicolumn{1}{l||}{0.53}                        & \multicolumn{1}{l|}{141}                           & \multicolumn{1}{l||}{0.8}                           & \multicolumn{1}{l|}{206.4}                        & \multicolumn{1}{l||}{0.88}                         & \multicolumn{1}{l|}{209.1}                         & \multicolumn{1}{l||}{0.89}                         & \multicolumn{1}{l|}{205.8}                         & \multicolumn{1}{l||}{0.95}                        & \multicolumn{1}{l|}{219.9}                        & \multicolumn{1}{l||}{0.96}                        & \multicolumn{1}{l|}{229.3}                        & \multicolumn{1}{l||}{0.96}                         & \multicolumn{1}{l|}{252}                           & \multicolumn{1}{l||}{0.97}                        & \multicolumn{1}{l|}{276.8}                     & \multicolumn{1}{l||}{0.98}                     \\ \cline{2-20} 
\rowcolor[HTML]{C0C0C0} 
{\cellcolor[HTML]{656565}{\color[HTML]{FFFFFF} LAL}}          & \begin{tabular}[c]{@{}c@{}}TAR$|$TTR\end{tabular}                               & \multicolumn{1}{l|}{\cellcolor[HTML]{C0C0C0}0.3}   & \multicolumn{1}{l||}{\cellcolor[HTML]{C0C0C0}0.54} & \multicolumn{1}{l|}{\cellcolor[HTML]{C0C0C0}0.5}   & \multicolumn{1}{l||}{\cellcolor[HTML]{C0C0C0}0.81}  & \multicolumn{1}{l|}{\cellcolor[HTML]{C0C0C0}0.74} & \multicolumn{1}{l||}{\cellcolor[HTML]{C0C0C0}0.89} & \multicolumn{1}{l|}{\cellcolor[HTML]{C0C0C0}0.76}  & \multicolumn{1}{l||}{\cellcolor[HTML]{C0C0C0}0.9}  & \multicolumn{1}{l|}{\cellcolor[HTML]{C0C0C0}0.74}  & \multicolumn{1}{l||}{\cellcolor[HTML]{C0C0C0}0.96} & \multicolumn{1}{l|}{\cellcolor[HTML]{C0C0C0}0.79} & \multicolumn{1}{l||}{\cellcolor[HTML]{C0C0C0}0.97} & \multicolumn{1}{l|}{\cellcolor[HTML]{C0C0C0}0.82} & \multicolumn{1}{l||}{\cellcolor[HTML]{C0C0C0}0.98}  & \multicolumn{1}{l|}{\cellcolor[HTML]{C0C0C0}0.91}  & \multicolumn{1}{l||}{\cellcolor[HTML]{C0C0C0}0.99} & \multicolumn{1}{l|}{\cellcolor[HTML]{C0C0C0}-} & \multicolumn{1}{l||}{\cellcolor[HTML]{C0C0C0}-} \\ \hline
\end{tabular}
\end{table*}

\begin{table*}[]
\tiny	
\caption{The results on VPN-nonVPN dataset (ISCXVPN2016).}
\label{tbl:vpn}
\begin{tabular}{|c|c||c|c||c|c||c|c||c|c||c|c||c|c||c|c||c|c||c|c||}
\hline
\rowcolor[HTML]{656565} 
{\color[HTML]{FFFFFF} \begin{tabular}[c]{@{}c@{}}Query\\  Strategy\end{tabular}} & {\color[HTML]{FFFFFF} \begin{tabular}[c]{@{}c@{}}Evaluated \\ Parameters\end{tabular}} & \multicolumn{2}{c||}{\cellcolor[HTML]{656565}{\color[HTML]{FFFFFF} 0.5\%}}                               & \multicolumn{2}{c||}{\cellcolor[HTML]{656565}{\color[HTML]{FFFFFF} 1\%}}                               & \multicolumn{2}{c||}{\cellcolor[HTML]{656565}{\color[HTML]{FFFFFF} 2\%}}                                & \multicolumn{2}{c||}{\cellcolor[HTML]{656565}{\color[HTML]{FFFFFF} 4\%}}                                & \multicolumn{2}{c||}{\cellcolor[HTML]{656565}{\color[HTML]{FFFFFF} 8\%}}                                & \multicolumn{2}{c||}{\cellcolor[HTML]{656565}{\color[HTML]{FFFFFF} 16\%}}                               & \multicolumn{2}{c||}{\cellcolor[HTML]{656565}{\color[HTML]{FFFFFF} 32\%}}                               & \multicolumn{2}{c||}{\cellcolor[HTML]{656565}{\color[HTML]{FFFFFF} 64\%}}                               & \multicolumn{2}{c||}{\cellcolor[HTML]{656565}{\color[HTML]{FFFFFF} Full}}                \\  \hline 
\cellcolor[HTML]{656565}{\color[HTML]{FFFFFF} }                                  & Time$|$ACC                                                                               & 7.2                                                & 0.49                                               & 12.4                                              & 0.52                                             & 14.1                                               & 0.57                                             & 25.7                                              & 0.58                                              & 38.4                                              & 0.61                                               & 60.4                                              & 0.62                                               & 101.9                                             & 0.62                                              & 162.1                                             & 0.61                                              & 218.9                                          & 0.62                                          \\ \cline{2-20} 
\rowcolor[HTML]{C0C0C0} 
{\cellcolor[HTML]{656565}{\color[HTML]{FFFFFF} QBC}}             & \begin{tabular}[c]{@{}c@{}}TAR$|$TTR\end{tabular}                               & 0.03                                               & 0.78                                               & 0.05                                              & 0.83                                              & 0.06                                               & 0.91                                             & 0.11                                              & 0.93                                              & 0.17                                              & 0.97                                              & 0.27                                              & 1                                                  & 0.46                                              & 1                                                  & 0.74                                              & 0.98                                               & -                                              & -                                              \\ \hline \hline
\cellcolor[HTML]{656565}{\color[HTML]{FFFFFF} }                                  & Time$|$ACC                                                                               & 2.2                                                & 0.46                                              & 4.1                                               & 0.50                                              & 5.6                                                & 0.55                                             & 12.5                                              & 0.58                                              & 21.4                                              & 0.57                                              & 40.1                                              & 0.57                                              & 76.02                                             & 0.59                                              & 132.5                                             & 0.62                                              & 178.4                                          & 0.63                                           \\ \cline{2-20} 
\rowcolor[HTML]{C0C0C0} 
{\cellcolor[HTML]{656565}{\color[HTML]{FFFFFF} UNC}}            & \begin{tabular}[c]{@{}c@{}}TAR$|$TTR\end{tabular}                               & 0.012                                              & 0.74                                              & 0.023                                             & 0.80                                              & 0.031                                              & 0.88                                              & 0.07                                              & 0.93                                               & 0.12                                              & 0.90                                               & 0.22                                              & 0.93                                               & 0.42                                              & 0.94                                               & 0.74                                              & 0.99                                               & -                                              & -                                              \\ \hline \hline
\cellcolor[HTML]{656565}{\color[HTML]{FFFFFF} }                                  & Time$|$ACC                                                                               & 1.9                                                & 0.493                                              & 3.6                                               & 0.52                                              & 4.5                                                & 0.56                                              & 6.2                                               & 0.57                                               & 9.5                                               & 0.58                                               & 28.4                                              & 0.59                                              & 53.9                                              & 0.60                                               & 84.8                                              & 0.62                                                & 115.6                                          & 0.62                                           \\ \cline{2-20} 
\rowcolor[HTML]{C0C0C0} 
{\cellcolor[HTML]{656565}{\color[HTML]{FFFFFF} Random}}         & \begin{tabular}[c]{@{}c@{}}TAR$|$TTR\end{tabular}                               & 0.016                                              & 0.79                                               & 0.031                                             & 0.83                                              & 0.039                                              & 0.9                                               & 0.05                                             & 0.91                                              & 0.08                                              & 0.94                                               & 0.24                                              & 0.95                                               & 0.46                                              & 0.97                                               & 0.73                                             & 1                                                  & -                                              & -                                              \\ \hline \hline
\cellcolor[HTML]{656565}{\color[HTML]{FFFFFF} }                                  & Time$|$ACC                                                                               & \multicolumn{1}{l|}{3.7}                           & \multicolumn{1}{l||}{0.47}                          & \multicolumn{1}{l|}{7}                            & \multicolumn{1}{l||}{0.49}                        & \multicolumn{1}{l|}{10.8}                          & \multicolumn{1}{l||}{0.54}                        & \multicolumn{1}{l|}{18.4}                         & \multicolumn{1}{l||}{0.56}                         & \multicolumn{1}{l|}{34.2}                         & \multicolumn{1}{l||}{0.56}                         & \multicolumn{1}{l|}{77.8}                         & \multicolumn{1}{l||}{0.57}                         & \multicolumn{1}{l|}{161.8}                        & \multicolumn{1}{l||}{0.60}                          & \multicolumn{1}{l|}{338.1}                        & \multicolumn{1}{l||}{0.61}                         & \multicolumn{1}{l|}{570}                       & \multicolumn{1}{l||}{0.61}                     \\ \cline{2-20} 
\rowcolor[HTML]{C0C0C0} 
{\cellcolor[HTML]{656565}{\color[HTML]{FFFFFF} Density}}         & \begin{tabular}[c]{@{}c@{}}TAR$|$TTR\end{tabular}                               & \multicolumn{1}{l|}{\cellcolor[HTML]{C0C0C0}0.006} & \multicolumn{1}{l||}{\cellcolor[HTML]{C0C0C0}0.77}  & \multicolumn{1}{l|}{\cellcolor[HTML]{C0C0C0}0.01} & \multicolumn{1}{l||}{\cellcolor[HTML]{C0C0C0}0.81} & \multicolumn{1}{l|}{\cellcolor[HTML]{C0C0C0}0.018} & \multicolumn{1}{l||}{\cellcolor[HTML]{C0C0C0}0.88} & \multicolumn{1}{l|}{\cellcolor[HTML]{C0C0C0}0.03} & \multicolumn{1}{l||}{\cellcolor[HTML]{C0C0C0}0.91}  & \multicolumn{1}{l|}{\cellcolor[HTML]{C0C0C0}0.06} & \multicolumn{1}{l||}{\cellcolor[HTML]{C0C0C0}0.91} & \multicolumn{1}{l|}{\cellcolor[HTML]{C0C0C0}0.13} & \multicolumn{1}{l||}{\cellcolor[HTML]{C0C0C0}0.93}  & \multicolumn{1}{l|}{\cellcolor[HTML]{C0C0C0}0.28} & \multicolumn{1}{l||}{\cellcolor[HTML]{C0C0C0}0.98} & \multicolumn{1}{l|}{\cellcolor[HTML]{C0C0C0}0.59} & \multicolumn{1}{l||}{\cellcolor[HTML]{C0C0C0}1}     & \multicolumn{1}{l|}{\cellcolor[HTML]{C0C0C0}-} & \multicolumn{1}{l||}{\cellcolor[HTML]{C0C0C0}-} \\ \hline \hline
\cellcolor[HTML]{656565}{\color[HTML]{FFFFFF} }                                  & Time$|$ACC                                                                               & \multicolumn{1}{l|}{141.9}                         & \multicolumn{1}{l||}{0.47}                          & \multicolumn{1}{l|}{205.3}                        & \multicolumn{1}{l||}{0.57}                         & \multicolumn{1}{l|}{204.9}                         & \multicolumn{1}{l||}{0.54}                        & \multicolumn{1}{l|}{204.8}                        & \multicolumn{1}{l||}{0.56}                          & \multicolumn{1}{l|}{209.3}                        & \multicolumn{1}{l||}{0.54}                          & \multicolumn{1}{l|}{230.5}                        & \multicolumn{1}{l||}{0.54}                         & \multicolumn{1}{l|}{253.9}                        & \multicolumn{1}{l||}{0.59}                         & \multicolumn{1}{l|}{283.1}                        & \multicolumn{1}{l||}{0.6}                           & \multicolumn{1}{l|}{316.3}                     & \multicolumn{1}{l||}{0.61}                     \\ \cline{2-20} 
\rowcolor[HTML]{C0C0C0} 
{\cellcolor[HTML]{656565}{\color[HTML]{FFFFFF} LAL}}             & \begin{tabular}[c]{@{}c@{}}TAR$|$TTR\end{tabular}                               & \multicolumn{1}{l|}{\cellcolor[HTML]{C0C0C0}0.44}  & \multicolumn{1}{l||}{\cellcolor[HTML]{C0C0C0}0.76} & \multicolumn{1}{l|}{\cellcolor[HTML]{C0C0C0}0.64} & \multicolumn{1}{l||}{\cellcolor[HTML]{C0C0C0}0.92} & \multicolumn{1}{l|}{\cellcolor[HTML]{C0C0C0}0.64}  & \multicolumn{1}{l||}{\cellcolor[HTML]{C0C0C0}0.88} & \multicolumn{1}{l|}{\cellcolor[HTML]{C0C0C0}0.64} & \multicolumn{1}{l||}{\cellcolor[HTML]{C0C0C0}0.91} & \multicolumn{1}{l|}{\cellcolor[HTML]{C0C0C0}0.66} & \multicolumn{1}{l||}{\cellcolor[HTML]{C0C0C0}0.88}  & \multicolumn{1}{l|}{\cellcolor[HTML]{C0C0C0}0.72} & \multicolumn{1}{l||}{\cellcolor[HTML]{C0C0C0}0.88} & \multicolumn{1}{l|}{\cellcolor[HTML]{C0C0C0}0.8}  & \multicolumn{1}{l||}{\cellcolor[HTML]{C0C0C0}0.97}  & \multicolumn{1}{l|}{\cellcolor[HTML]{C0C0C0}0.89} & \multicolumn{1}{l||}{\cellcolor[HTML]{C0C0C0}0.97} & \multicolumn{1}{l|}{\cellcolor[HTML]{C0C0C0}-} & \multicolumn{1}{l||}{\cellcolor[HTML]{C0C0C0}-} \\ \hline 
\end{tabular}
\end{table*}

\begin{table*}[]
\tiny	
\caption{The results on Tor No-Tor dataset.}
\label{tbl:tornotor}
\begin{tabular}{|c|c||c|c||c|c||c|c||c|c||c|c||c|c||c|c||c|c||c|c||}
\hline
\rowcolor[HTML]{656565} 
{\color[HTML]{FFFFFF} \begin{tabular}[c]{@{}c@{}}Query\\  Strategy\end{tabular}} & {\color[HTML]{FFFFFF} \begin{tabular}[c]{@{}c@{}}Evaluated\\  Parameters\end{tabular}}                                 & \multicolumn{2}{c||}{\cellcolor[HTML]{656565}{\color[HTML]{FFFFFF} 0.5\%}}                           & \multicolumn{2}{c||}{\cellcolor[HTML]{656565}{\color[HTML]{FFFFFF} 1\%}}                             & \multicolumn{2}{c||}{\cellcolor[HTML]{656565}{\color[HTML]{FFFFFF} 2\%}}                                & \multicolumn{2}{c||}{\cellcolor[HTML]{656565}{\color[HTML]{FFFFFF} 4\%}}                             & \multicolumn{2}{c||}{\cellcolor[HTML]{656565}{\color[HTML]{FFFFFF} 8\%}}                            & \multicolumn{2}{c||}{\cellcolor[HTML]{656565}{\color[HTML]{FFFFFF} 16\%}}                           & \multicolumn{2}{c||}{\cellcolor[HTML]{656565}{\color[HTML]{FFFFFF} 32\%}}                           & \multicolumn{2}{c||}{\cellcolor[HTML]{656565}{\color[HTML]{FFFFFF} 64\%}}                           & \multicolumn{2}{c||}{\cellcolor[HTML]{656565}{\color[HTML]{FFFFFF} Full}}                        \\ \hline
\cellcolor[HTML]{656565}{\color[HTML]{FFFFFF} }                                  & Time$|$ACC                                                 & 16.10                                              & 0.8                                            & 25.1                                               & 0.83                                           & 41.53                                              & 0.84                                             & 65.56                                              & 0.84                                          & 110.7                                             & 0.83                                          & 175                                               & 0.82                                          & 286.1                                             & 0.83                                          & 510                                               & 0.83                                          & 771.3                                          & 0.83                                          \\ \cline{2-20} 
\rowcolor[HTML]{C0C0C0} 
{\cellcolor[HTML]{656565}{\color[HTML]{FFFFFF} QBC}}            & \begin{tabular}[c]{@{}c@{}}TAR$|$TTR\end{tabular} & 0.02                                               & 0.95                                           & 0.03                                               & 1                                              & 0.05                                               & {\color[HTML]{000000} 1}                          & 0.08                                               & 1                                              & 0.14                                              & 1                                              & 0.22                                              & 0.98                                           & 0.37                                              & 1                                              & 0.66                                              & 1                                              & -                                              & -                                              \\ \hline \hline
\cellcolor[HTML]{656565}{\color[HTML]{FFFFFF} }                                  & Time$|$ACC                                                 & 6.4                                                & 0.82                                          & 12.2                                               & 0.83                                          & 25.2                                               & 0.82                                             & 45.7                                               & 0.83                                          & 87.11                                             & 0.83                                          & 147.5                                             & 0.83                                          & 242.1                                             & 0.83                                          & 420.5                                             & 0.83                                          & 636.2                                          & 0.82                                          \\ \cline{2-20} 
\rowcolor[HTML]{C0C0C0} 
{\cellcolor[HTML]{656565}{\color[HTML]{FFFFFF} UNC}}            & \begin{tabular}[c]{@{}c@{}}TAR$|$TTR\end{tabular} & 0.01                                               & 1                                              & 0.019                                              & 1                                              & 0.03                                               & 1                                                 & 0.07                                               & 1                                              & 0.13                                              & 1                                              & 0.23                                              & 1                                              & 0.38                                              & 1                                              & 0.66                                              & 1                                              & -                                              & -                                              \\ \hline \hline
\cellcolor[HTML]{656565}{\color[HTML]{FFFFFF} }                                  & Time$|$ACC                                                 & 4.8                                                & 0.83                                           & 6.7                                                & 0.83                                           & 10.3                                               & 0.84                                              & 31.84                                              & 0.81                                          & 59.1                                              & 0.82                                          & 90.5                                              & 0.82                                          & 151                                               & 0.82                                          & 273.7                                             & 0.83                                          & 417.2                                          & 0.82                                           \\ \cline{2-20} 
\rowcolor[HTML]{C0C0C0} 
{\cellcolor[HTML]{656565}{\color[HTML]{FFFFFF} Random}}          & \begin{tabular}[c]{@{}c@{}}TAR$|$TTR\end{tabular} & 0.011                                              & 1                                              & 0.016                                              & 1                                              & 0.024                                              & 1                                                 & 0.076                                              & 0.99                                           & 0.14                                              & 1                                              & 0.21                                              & 1                                              & 0.36                                              & 1                                              & 0.65                                              & 1                                              & -                                              & -                                              \\ \hline \hline
\cellcolor[HTML]{656565}{\color[HTML]{FFFFFF} }                                  & Time$|$ACC                                                 & \multicolumn{1}{l|}{11.39}                         & \multicolumn{1}{l||}{0.83}                      & \multicolumn{1}{l|}{19.44}                         & \multicolumn{1}{l||}{0.84}                      & \multicolumn{1}{l|}{36.2}                          & \multicolumn{1}{l||}{0.80}                        & \multicolumn{1}{l|}{83.35}                         & \multicolumn{1}{l||}{0.82}                     & \multicolumn{1}{l|}{170.4}                        & \multicolumn{1}{l||}{0.83}                     & \multicolumn{1}{l|}{345.9}                        & \multicolumn{1}{l||}{0.82}                     & \multicolumn{1}{l|}{773.2}                        & \multicolumn{1}{l||}{0.827}                     & \multicolumn{1}{l|}{2064.8}                       & \multicolumn{1}{l||}{0.82}                     & \multicolumn{1}{l|}{3973.4}                    & \multicolumn{1}{l||}{0.82}                      \\ \cline{2-20} 
\rowcolor[HTML]{C0C0C0} 
{\cellcolor[HTML]{656565}{\color[HTML]{FFFFFF} Density}}        & \begin{tabular}[c]{@{}c@{}}TAR$|$TTR\end{tabular} & \multicolumn{1}{l|}{\cellcolor[HTML]{C0C0C0}0.002} & \multicolumn{1}{l||}{\cellcolor[HTML]{C0C0C0}1} & \multicolumn{1}{l|}{\cellcolor[HTML]{C0C0C0}0.004} & \multicolumn{1}{l||}{\cellcolor[HTML]{C0C0C0}1} & \multicolumn{1}{l|}{\cellcolor[HTML]{C0C0C0}0.009} & \multicolumn{1}{l||}{\cellcolor[HTML]{C0C0C0}0.97} & \multicolumn{1}{l|}{\cellcolor[HTML]{C0C0C0}0.02}  & \multicolumn{1}{l||}{\cellcolor[HTML]{C0C0C0}1} & \multicolumn{1}{l|}{\cellcolor[HTML]{C0C0C0}0.04} & \multicolumn{1}{l||}{\cellcolor[HTML]{C0C0C0}1} & \multicolumn{1}{l|}{\cellcolor[HTML]{C0C0C0}0.08} & \multicolumn{1}{l||}{\cellcolor[HTML]{C0C0C0}1} & \multicolumn{1}{l|}{\cellcolor[HTML]{C0C0C0}0.19} & \multicolumn{1}{l||}{\cellcolor[HTML]{C0C0C0}1} & \multicolumn{1}{l|}{\cellcolor[HTML]{C0C0C0}0.51} & \multicolumn{1}{l||}{\cellcolor[HTML]{C0C0C0}1} & \multicolumn{1}{l|}{\cellcolor[HTML]{C0C0C0}-} & \multicolumn{1}{l||}{\cellcolor[HTML]{C0C0C0}-} \\ \hline \hline
\cellcolor[HTML]{656565}{\color[HTML]{FFFFFF} }                                  & Time$|$ACC                                                 & \multicolumn{1}{l|}{204.9}                        & \multicolumn{1}{l||}{0.83}                     & \multicolumn{1}{l|}{203.1}                         & \multicolumn{1}{l||}{0.82}                     & \multicolumn{1}{l|}{207.5}                         & \multicolumn{1}{l||}{0.81}                         & \multicolumn{1}{l|}{229.9}                         & \multicolumn{1}{l||}{0.82}                     & \multicolumn{1}{l|}{254.1}                        & \multicolumn{1}{l||}{0.82}                     & \multicolumn{1}{l|}{289.7}                        & \multicolumn{1}{l||}{0.83}                     & \multicolumn{1}{l|}{348.4}                        & \multicolumn{1}{l||}{0.82}                     & \multicolumn{1}{l|}{471.9}                        & \multicolumn{1}{l||}{0.83}                      & \multicolumn{1}{l|}{642.2}                     & \multicolumn{1}{l||}{0.82}                     \\ \cline{2-20} 
\rowcolor[HTML]{C0C0C0} 
{\cellcolor[HTML]{656565}{\color[HTML]{FFFFFF} LAL}}          & \begin{tabular}[c]{@{}c@{}}TAR$|$TTR\end{tabular} & \multicolumn{1}{l|}{\cellcolor[HTML]{C0C0C0}0.31}  & \multicolumn{1}{l||}{\cellcolor[HTML]{C0C0C0}1} & \multicolumn{1}{l|}{\cellcolor[HTML]{C0C0C0}0.31}  & \multicolumn{1}{l||}{\cellcolor[HTML]{C0C0C0}1} & \multicolumn{1}{l|}{\cellcolor[HTML]{C0C0C0}0.32}  & \multicolumn{1}{l||}{\cellcolor[HTML]{C0C0C0}0.98} & \multicolumn{1}{l|}{\cellcolor[HTML]{C0C0C0}0.35} & \multicolumn{1}{l||}{\cellcolor[HTML]{C0C0C0}1} & \multicolumn{1}{l|}{\cellcolor[HTML]{C0C0C0}0.39} & \multicolumn{1}{l||}{\cellcolor[HTML]{C0C0C0}1} & \multicolumn{1}{l|}{\cellcolor[HTML]{C0C0C0}0.45} & \multicolumn{1}{l||}{\cellcolor[HTML]{C0C0C0}1} & \multicolumn{1}{l|}{\cellcolor[HTML]{C0C0C0}0.54} & \multicolumn{1}{l||}{\cellcolor[HTML]{C0C0C0}1} & \multicolumn{1}{l|}{\cellcolor[HTML]{C0C0C0}0.73} & \multicolumn{1}{l||}{\cellcolor[HTML]{C0C0C0}1} & \multicolumn{1}{l|}{\cellcolor[HTML]{C0C0C0}-} & \multicolumn{1}{l||}{\cellcolor[HTML]{C0C0C0}-} \\ \hline
\end{tabular}
\end{table*}


\section{Core Challenges and considerations}
\label{sec:challenges}
Although \gls{al} can achieve a reasonable level of accuracy using much less labeled examples than traditional ML algorithms, there are still some unsolved problems.
In this section, we discuss those challenges and identify the relevant open issues and considerations.

\subsection{Challenges}
\textcolor{black}{The challenges of using \gls{al} for \gls{ntc} are listed as below. Table \ref{tbl:challenges} summarizes the challenges and some literature that try to \textcolor{black}{address those challenges.}}
\begin{itemize}
\item \textbf{Noisy annotation:} In conventional ML, the labels are often assumed to be the ground truth without noise; However, when the instances are annotated by humans or machines that are prone to errors, additional considerations must be made \cite{artstein2008inter}. The existence of noisy labels is problematic for the uncertainty sampling strategy, due to the fact that this strategy is intrinsically noise-seeking \cite{GARCIA2015108}. In networking, noises can be generated deliberately or accidentally. Intruders can generate packets to mislead NTC systems. For example, in \cite{perdisci2006misleading} the problem of noise in worm signatures extraction algorithms has been investigated. In these algorithms, deliberate noise can prevent from building reliable and useful worm signatures, and consequently decline the performance of \glspl{ids}. Noises can also be generated accidentally \textcolor{black}{due to} human or automated process labeling mistakes. In \cite{donmez2008proactive}, Donmez et al. proposed proactive learning to deal with noisy labels as it uses a decision-theoretic method in order to jointly choose the optimal oracle and instance. 
In \cite{nguyen2004active} and \cite{donmez2007dual}, choosing the most informative instances and combining density-weighted uncertainty sampling and standard uncertainty sampling are proposed respectively to address the noisy annotation issue.


\item \textbf{Stopping criteria:} \textcolor{black}{The stopping criteria determine when to stop querying unlabeled instances, such as reaching the desired accuracy, running time, or limiting the number of queries, to name a few.
As \gls{ntc} is sensitive in both time and accuracy aspects, providing efficient stopping criteria is necessary.
As an example, in case of security support for network traffic, accuracy has a higher priority, but multimedia streams are mainly time-sensitive. 
In case of latter, the adoption of right stopping criteria can reduce the time complexity of \gls{al} techniques for re-training the model.}
\textcolor{black}{Moreover, many studies on \gls{al} have adopted the convergence of the error rate on a set of unseen data samples as the stopping criterion.} Nevertheless, providing a new annotated unseen dataset is in direct contradiction to the goal of \gls{al}, due to the fact that the purpose is to reduce the need for annotation. In addition, if the unseen dataset contains a small number of instances, there is no guarantee that this dataset will be representative of the future instances. To alleviate this challenge, the authors in \cite{hakkani2006active, altschuler2019stopping, bloodgood2014method, temerinac2015deciding} proposed different approaches for stopping criteria based on, \eg, stabilization of forecast over different iterations, batch-wise learning, etc.



\item \textbf{Outliers in the data:} In a canonical definition, an outlier is a data example that considerably differs from other data points in a dataset, which may occur by chance or due to an experimental error resulting in inefficiency to analyze the overall behavior of the dataset. Labeling outlier instances by a human/machine annotator may not help in the prediction of the labels of unlabeled instances, because these instances are likely to be representative of a small number of instances. 
\textcolor{black}{{In \gls{ntc} data, outliers can appear due to networking problems such as inefficient data forwarding models and network congestion.} There are two major approaches to deal with outliers, including correcting and removing outlier data. \textcolor{black}{Methods such as Z-Score, smoothing techniques, and interquartile range (IQR)-Score are the common methods for removing outlier data.} 
Apart from the dealing method, the first step towards handling outliers is outliers detection. \textcolor{black}{There are many methods for outliers detection.} Model-based method is among the simplest methods, in which one constructs a model for all instances, and then defines outliers as instances having significant deviations from the constructed profiles \cite{yang2009outlier}. The proximity-based method is another \textcolor{black}{popular method, which 
can be realized}through algorithms such as K-th nearest neighbor, the ratio of local reachability density, and log transformation \cite{ramaswamy2000efficient,breunig2000lof,warner2008applied}. Generally, solutions both for correcting and removing outliers can be applied to \gls{ntc} techniques. Practically, outlier correction techniques are more complicated as they are rather time- and resource-demanding. The main advantage of outlier correction techniques is that they can be useful in case of data scarcity as availability of more data can improve training in \gls{ml} techniques.} On the contrary, outlier \textcolor{black}{removal is more} efficient in terms of speed. Therefore, based on the target quality requirements such as resource consumption or delay-sensitivity we can choose the appropriate solution for dealing with outliers.


\item \textbf{Drifting distribution:} Most existing \gls{al} algorithms work on the assumption that the data is \gls{iid}; However, the selection of instances based on a non-random distribution, which shifts over time, may lead to divergence between the prior distribution of instances and the current one \cite{hakkani2006active, yang2011active, vzliobaite2013active}. Such divergence can affect the performance of the learning task (\ie, classification or regression), since \gls{iid} assumption is no longer valid. In \gls{ntc}, drifting distribution in the gathered data from the network can appear because of different events, \eg network congestion, inefficient queuing models in middle nodes, and inefficient packet inspection model.

\item \textbf{Instance selection criteria:} 
In query strategies, criteria for selecting the instances are very important for an \gls{al} system as the system uses the selected instances like a training set.
\textcolor{black}{Despite having several query strategies proposed, such as those described in Section \ref{sec:al}, there is the} lack of research on the role of the annotator (human or machine) in the labeling process. In other words, the annotator plays a passive role in providing a label to an instance, and the lack of understanding about the importance of querying is apparent. 
In addition, different query strategies can be employed based on the volume of existing samples and the essence of the network traffic. Based on the \textit{Theory of Network} challenge, there is no specific policy to use query strategies as the behaviors of networks are different.

\item \textbf{Labeling method:} Although Section \ref{sec:performance} shows that \gls{al} techniques are efficient for \gls{ntc}, especially improving the training time and dealing with the shortage of data samples, the main drawback of using \gls{al} is the lack of a concrete labelling technique. Due to the \textit{Concept Drift} challenge, the model should be retrained frequently. On the other hand, as mentioned above, labelling the selected instances can be performed by human or machine, but in real-world applications of \gls{ntc}, using the human-labour is almost impossible due to the slowness and inefficiency issues. \textcolor{black}{Considering that in most real-world applications of networking, data is unlabeled or semi-labeled \cite{alsheikh2016mobile}, a machine-based labeling method is highly needed to perform labelling at the speed of the network with a high accuracy in labeling.
The authors in \cite{tu2020better} have proposed an interesting idea of combining human and artificial expertise for labeling commits in software development, which has the penitential to be applied in NTC and meet the mentioned criteria.}

\end{itemize}
\begin{table*}[]
\scriptsize
\caption{\textcolor{black}{Active learning core challenges and solutions. }}
\label{tbl:challenges}
\begin{tabular}{|l|l|}

\hline
\textbf{Challenges} &\textbf{Proposed solution (s) in literature} \\ \hline
 Noisy annotations & \textbf{(1)} Proactive learning \cite{donmez2008proactive}, \textbf{(2)} density-weighted uncertainty sampling \cite{nguyen2004active}, \textbf{(3)} standard uncertainty sampling \cite{donmez2007dual} \\ \hline

 Stopping criteria & \textbf{(1)} Stabilizing predictions \cite{bloodgood2014method}, \textbf{(2)} batch-wise learning \cite{temerinac2015deciding} \\ \hline

 Outliers in data & \textbf{(1)} Model-based method \cite{yang2009outlier}, \textbf{(2)} LOF algorithm \cite{breunig2000lof} \\ \hline

 Drifting distribution & \textbf{(1)} AL strategies based on uncertainty, randomization of the search space, and dynamic allocation of labeling attempts steadily \cite{vzliobaite2013active} \\ \hline

 Instance selection criteria & \textbf{(1)} Greedy AL strategy \cite{dasgupta2005analysis}, \textbf{(2)} expected model output changes strategy \cite{freytag2014selecting}, \textbf{(3)} instance weighting-based AL strategy \cite{bouguelia2016adaptive} \\ \hline

 Labeling method & \textbf{(1)} Human+artificial labelling method \cite{tu2020better} \\ \hline





\end{tabular}
\end{table*}

\subsection{Considerations and open issues}
\textcolor{black}{There are some considerations and open issues that should be studied to use \gls{al} models in \gls{ntc} as \textcolor{black}{discussed} below. }
\begin{itemize}
\item \textbf{The existing gap between deep learning and \gls{al} techniques:}
\textcolor{black}{Deep learning models are being deployed in a diverse range of real-world applications, such as self-driving cars, anomaly detection, wearables, \gls{ntma} \cite{abbasi2021deep} and healthcare. However, to achieve} the full potential of deep learning in such applications, the lack of labeled data is one of the major barriers. Recent works in the context of \gls{al} focus on proposing new/improved algorithms, and a few of them deal with the problem of intelligent data collection \cite{2020arXiv200900236R}. From a data science perspective, it is desirable to fill this gap. We believe that conducting further research on deep \gls{al} \cite{pimentel2018deep} and human-in-the-loop \gls{ml} \cite{yang2018active} will narrow the gap. In these branches of \gls{ml}, intelligent data collection is a step of the \gls{ml} process. Human-in-the-loop \gls{ml} can further facilitate the process of labeling difficult/new instances that a machine annotator cannot handle. 
 
 \item \textbf{NTMA applications in which \gls{al} is still missing:} 
 As mentioned above, many \gls{al} works deploy AL for security purposes, such as intrusion detection and malware detection. However, there is still a long list of \gls{al} applications, such as traffic forecasting, fault management, QoS management, and routing, in which AL has not been employed. For instance, to our best knowledge, no studies have been conducted on the coupling of network traffic prediction with AL. This is maybe due to the dynamic nature of traffic prediction tasks and difficulty of working with streaming data. The traffic load on communication networks changes overtime or in different situations. Indeed, the distribution of the traffic load may face the problem of drift distribution.
 
 \item \textbf{Lack of appropriate public network traffic datasets:} \gls{ntc} involves the categorization of network traffic into several traffic classes, such as WWW, mail, attack, \gls{p2p}, and multimedia. However, as mentioned above, network security becomes a dominant application of \gls{al}-based traffic classification. For example, we have found only two papers that have used \gls{al} for \gls{p2p} traffic classification \cite{dai2012peer}, \cite{liu2015active}. This is because there are a few relevant and public non-security related traffic datasets. Moreover, the datasets used in these papers are often not updated. The ever-growing popularity in using encryption protocols and the number of Internet-based applications calls for up-to-date non-security related traffic datasets to employ for \gls{ml} techniques. \textcolor{black}{Although Theory of Network indicates that benchmark datasets cannot be used in all networks, having the datasets can help at least evaluate the performance of \gls{ml} techniques, \eg \gls{al} in \gls{ntc}.}
 
 \item \textbf{Uncertainty in the performance of \gls{al} based on \textit{Theory of Network}:} Although the performance of \gls{al} is evaluated based on benchmark datasets, there is no guarantee to have such performance in each network separately. Different reasons, \eg weak samples, \gls{ml} techniques, the complexity of network traffics, etc. can affect the performance of the \gls{al} in each network. Over-fitting can be a considerable challenge in different networks regarding the use of \gls{al}. Although in this study we evaluate the performance of \gls{al} in a general manner, its performance should be \textcolor{black}{studied in the real-world target network}. 
 
 \item \textcolor{black}{\textbf{Resource-constrained networks:}
 Most \gls{ml} algorithms including \gls{al} are designed to run on resource-rich devices, but in real-world applications, many devices are resource-poor, \eg \gls{iot}, Edge and Fog devices. Thus, it is needed to optimize \gls{ml} techniques, \ie \gls{al} techniques, to be efficiently executable on such devices. 
\textcolor{black}{As a solution, different distributed learning techniques have been introduced to use resources in resource-constrained devices to run \gls{ml} algorithms in a distributed manner, \eg federated learning \cite{yang2019federated}.}}
\end{itemize}



\section{Conclusion}
\label{sec:conclusion}
In this article, active \gls{ml} for \gls{ntc} has been investigated. We first provided a background on \gls{ntc} and \gls{al}. The main traffic classification techniques and related issues are summarized, as well as the fundamental elements of \gls{al} are explained. Moreover, the core issues and challenges of \gls{al} are introduced. This article can also serve as a reference for how to use \gls{al} for \gls{ntc}. Using simulations, the impact of \gls{al} on the learning process was analyzed. Based on conducted investigations, the benefits of \gls{al} for \gls{ntc} are discussed throughout the paper. We showed that \gls{al} has an excellent opportunity to improve the performance of ML techniques to be used in \gls{ntc} as it can reduce the training time and the need of labeled data. As the main challenge of networking, network dynamicity enforces the ML techniques to be retrained frequently and \gls{al} helps retrain the model using a limited number of samples.

\bibliographystyle{IEEEtran}

\bibliography{mybib.bib}

\begin{thebibliography}{100}
\providecommand{\url}[1]{#1}
\csname url@samestyle\endcsname
\providecommand{\newblock}{\relax}
\providecommand{\bibinfo}[2]{#2}
\providecommand{\BIBentrySTDinterwordspacing}{\spaceskip=0pt\relax}
\providecommand{\BIBentryALTinterwordstretchfactor}{4}
\providecommand{\BIBentryALTinterwordspacing}{\spaceskip=\fontdimen2\font plus
\BIBentryALTinterwordstretchfactor\fontdimen3\font minus
  \fontdimen4\font\relax}
\providecommand{\BIBforeignlanguage}[2]{{%
\expandafter\ifx\csname l@#1\endcsname\relax
\typeout{** WARNING: IEEEtran.bst: No hyphenation pattern has been}%
\typeout{** loaded for the language `#1'. Using the pattern for}%
\typeout{** the default language instead.}%
\else
\language=\csname l@#1\endcsname
\fi
#2}}
\providecommand{\BIBdecl}{\relax}
\BIBdecl

\bibitem{wang2013internet}
Y.~Wang, Y.~Xiang, J.~Zhang, W.~Zhou, G.~Wei, and L.~T. Yang, ``Internet
  traffic classification using constrained clustering,'' \emph{IEEE
  transactions on parallel and distributed systems}, vol.~25, no.~11, pp.
  2932--2943, 2013.

\bibitem{ayoubi2018machine}
S.~Ayoubi, N.~Limam, M.~A. Salahuddin, N.~Shahriar, R.~Boutaba,
  F.~Estrada-Solano, and O.~M. Caicedo, ``Machine learning for cognitive
  network management,'' \emph{IEEE Communications Magazine}, vol.~56, no.~1,
  pp. 158--165, 2018.

\bibitem{casas2016big}
P.~Casas, A.~D'Alconzo, T.~Zseby, and M.~Mellia, ``Big-dama: big data analytics
  for network traffic monitoring and analysis,'' in \emph{Proceedings of the
  2016 workshop on Fostering Latin-American Research in Data Communication
  Networks}, 2016, pp. 1--3.

\bibitem{8789667}
A.~{D’Alconzo}, I.~{Drago}, A.~{Morichetta}, M.~{Mellia}, and P.~{Casas}, ``A
  survey on big data for network traffic monitoring and analysis,'' \emph{IEEE
  Transactions on Network and Service Management}, vol.~16, no.~3, pp.
  800--813, 2019.

\bibitem{stoyanova2020survey}
M.~Stoyanova, Y.~Nikoloudakis, S.~Panagiotakis, E.~Pallis, and E.~K. Markakis,
  ``A survey on the internet of things (iot) forensics: Challenges, approaches
  and open issues,'' \emph{IEEE Communications Surveys \& Tutorials}, 2020.

\bibitem{boutaba2018comprehensive}
R.~Boutaba, M.~A. Salahuddin, N.~Limam, S.~Ayoubi, N.~Shahriar,
  F.~Estrada-Solano, and O.~M. Caicedo, ``A comprehensive survey on machine
  learning for networking: evolution, applications and research
  opportunities,'' \emph{Journal of Internet Services and Applications},
  vol.~9, no.~1, p.~16, 2018.

\bibitem{tahaei2020rise}
H.~Tahaei, F.~Afifi, A.~Asemi, F.~Zaki, and N.~B. Anuar, ``The rise of traffic
  classification in iot networks: A survey,'' \emph{Journal of Network and
  Computer Applications}, vol. 154, p. 102538, 2020.

\bibitem{4738466}
T.~T.~T. {Nguyen} and G.~{Armitage}, ``A survey of techniques for internet
  traffic classification using machine learning,'' \emph{IEEE Communications
  Surveys Tutorials}, vol.~10, no.~4, pp. 56--76, 2008.

\bibitem{wang2019survey}
P.~Wang, X.~Chen, F.~Ye, and Z.~Sun, ``A survey of techniques for mobile
  service encrypted traffic classification using deep learning,'' \emph{IEEE
  Access}, vol.~7, pp. 54\,024--54\,033, 2019.

\bibitem{lopez2020iot}
M.~Lopez-Martin, B.~Carro, and A.~Sanchez-Esguevillas, ``Iot type-of-traffic
  forecasting method based on gradient boosting neural networks,'' \emph{Future
  Generation Computer Systems}, vol. 105, pp. 331--345, 2020.

\bibitem{8026581}
M.~{Lopez-Martin}, B.~{Carro}, A.~{Sanchez-Esguevillas}, and J.~{Lloret},
  ``Network traffic classifier with convolutional and recurrent neural networks
  for internet of things,'' \emph{IEEE Access}, vol.~5, pp. 18\,042--18\,050,
  2017.

\bibitem{alsheikh2016mobile}
M.~A. Alsheikh, D.~Niyato, S.~Lin, H.-P. Tan, and Z.~Han, ``Mobile big data
  analytics using deep learning and apache spark,'' \emph{IEEE network},
  vol.~30, no.~3, pp. 22--29, 2016.

\bibitem{tu2020better}
H.~Tu, Z.~Yu, and T.~Menzies, ``Better data labelling with emblem (and how that
  impacts defect prediction),'' \emph{IEEE Transactions on Software
  Engineering}, 2020.

\bibitem{settles1648active}
B.~Settles, ``Active learning literature survey univ. wisconsin-madison,
  madison, wi, 2009,'' \emph{CS Tech. Rep. 1648, Tech. Rep.}, 2009.

\bibitem{sun2019application}
Y.~Sun, M.~Peng, Y.~Zhou, Y.~Huang, and S.~Mao, ``Application of machine
  learning in wireless networks: Key techniques and open issues,'' \emph{IEEE
  Communications Surveys \& Tutorials}, vol.~21, no.~4, pp. 3072--3108, 2019.

\bibitem{chen2019artificial}
M.~Chen, U.~Challita, W.~Saad, C.~Yin, and M.~Debbah, ``Artificial neural
  networks-based machine learning for wireless networks: A tutorial,''
  \emph{IEEE Communications Surveys \& Tutorials}, vol.~21, no.~4, pp.
  3039--3071, 2019.

\bibitem{mao2018deep}
Q.~Mao, F.~Hu, and Q.~Hao, ``Deep learning for intelligent wireless networks: A
  comprehensive survey,'' \emph{IEEE Communications Surveys \& Tutorials},
  vol.~20, no.~4, pp. 2595--2621, 2018.

\bibitem{yau2012reinforcement}
K.-L.~A. Yau, P.~Komisarczuk, and P.~D. Teal, ``Reinforcement learning for
  context awareness and intelligence in wireless networks: Review, new features
  and open issues,'' \emph{Journal of Network and Computer Applications},
  vol.~35, no.~1, pp. 253--267, 2012.

\bibitem{xie2018survey}
J.~Xie, F.~R. Yu, T.~Huang, R.~Xie, J.~Liu, C.~Wang, and Y.~Liu, ``A survey of
  machine learning techniques applied to software defined networking (sdn):
  Research issues and challenges,'' \emph{IEEE Communications Surveys \&
  Tutorials}, vol.~21, no.~1, pp. 393--430, 2018.

\bibitem{gu2020machine}
R.~Gu, Z.~Yang, and Y.~Ji, ``Machine learning for intelligent optical networks:
  A comprehensive survey,'' \emph{Journal of Network and Computer
  Applications}, vol. 157, p. 102576, 2020.

\bibitem{singh2015performance}
H.~Singh, ``Performance analysis of unsupervised machine learning techniques
  for network traffic classification,'' in \emph{2015 Fifth International
  Conference on Advanced Computing \& Communication Technologies}.\hskip 1em
  plus 0.5em minus 0.4em\relax IEEE, 2015, pp. 401--404.

\bibitem{dainotti2012issues}
A.~Dainotti, A.~Pescape, and K.~C. Claffy, ``Issues and future directions in
  traffic classification,'' \emph{IEEE network}, vol.~26, no.~1, pp. 35--40,
  2012.

\bibitem{finsterbusch2013survey}
M.~Finsterbusch, C.~Richter, E.~Rocha, J.-A. Muller, and K.~Hanssgen, ``A
  survey of payload-based traffic classification approaches,'' \emph{IEEE
  Communications Surveys \& Tutorials}, vol.~16, no.~2, pp. 1135--1156, 2013.

\bibitem{velan2015survey}
P.~Velan, M.~{\v{C}}erm{\'a}k, P.~{\v{C}}eleda, and M.~Dra{\v{s}}ar, ``A survey
  of methods for encrypted traffic classification and analysis,''
  \emph{International Journal of Network Management}, vol.~25, no.~5, pp.
  355--374, 2015.

\bibitem{zhao2021network}
J.~Zhao, X.~Jing, Z.~Yan, and W.~Pedrycz, ``Network traffic classification for
  data fusion: A survey,'' \emph{Information Fusion}, vol.~72, pp. 22--47,
  2021.

\bibitem{nguyen2008survey}
T.~T. Nguyen and G.~Armitage, ``A survey of techniques for internet traffic
  classification using machine learning,'' \emph{IEEE communications surveys \&
  tutorials}, vol.~10, no.~4, pp. 56--76, 2008.

\bibitem{perera2017comparison}
P.~Perera, Y.-C. Tian, C.~Fidge, and W.~Kelly, ``A comparison of supervised
  machine learning algorithms for classification of communications network
  traffic,'' in \emph{International Conference on Neural Information
  Processing}.\hskip 1em plus 0.5em minus 0.4em\relax Springer, 2017, pp.
  445--454.

\bibitem{gomez2017ensemble}
S.~E. G{\'o}mez, B.~C. Mart{\'\i}nez, A.~J. S{\'a}nchez-Esguevillas, and L.~H.
  Callejo, ``Ensemble network traffic classification: Algorithm comparison and
  novel ensemble scheme proposal,'' \emph{Computer Networks}, vol. 127, pp.
  68--80, 2017.

\bibitem{pacheco2018towards}
F.~Pacheco, E.~Exposito, M.~Gineste, C.~Baudoin, and J.~Aguilar, ``Towards the
  deployment of machine learning solutions in network traffic classification: A
  systematic survey,'' \emph{IEEE Communications Surveys \& Tutorials},
  vol.~21, no.~2, pp. 1988--2014, 2018.

\bibitem{moore2005internet}
A.~W. Moore and D.~Zuev, ``Internet traffic classification using bayesian
  analysis techniques,'' in \emph{Proceedings of the 2005 ACM SIGMETRICS
  international conference on Measurement and modeling of computer systems},
  2005, pp. 50--60.

\bibitem{yuan2010svm}
R.~Yuan, Z.~Li, X.~Guan, and L.~Xu, ``An svm-based machine learning method for
  accurate internet traffic classification,'' \emph{Information Systems
  Frontiers}, vol.~12, no.~2, pp. 149--156, 2010.

\bibitem{rezaei2019deep}
S.~Rezaei and X.~Liu, ``Deep learning for encrypted traffic classification: An
  overview,'' \emph{IEEE communications magazine}, vol.~57, no.~5, pp. 76--81,
  2019.

\bibitem{ring2019survey}
M.~Ring, S.~Wunderlich, D.~Scheuring, D.~Landes, and A.~Hotho, ``A survey of
  network-based intrusion detection data sets,'' \emph{Computers \& Security},
  vol.~86, pp. 147--167, 2019.

\bibitem{kamiran2012data}
F.~Kamiran and T.~Calders, ``Data preprocessing techniques for classification
  without discrimination,'' \emph{Knowledge and Information Systems}, vol.~33,
  no.~1, pp. 1--33, 2012.

\bibitem{zheng2018feature}
A.~Zheng and A.~Casari, \emph{Feature engineering for machine learning:
  principles and techniques for data scientists}.\hskip 1em plus 0.5em minus
  0.4em\relax " O'Reilly Media, Inc.", 2018.

\bibitem{dong2018feature}
G.~Dong and H.~Liu, \emph{Feature engineering for machine learning and data
  analytics}.\hskip 1em plus 0.5em minus 0.4em\relax CRC Press, 2018.

\bibitem{sicker2007legal}
D.~C. Sicker, P.~Ohm, and D.~Grunwald, ``Legal issues surrounding monitoring
  during network research,'' in \emph{Proceedings of the 7th ACM SIGCOMM
  conference on Internet measurement}, 2007, pp. 141--148.

\bibitem{lopez2017network}
M.~Lopez-Martin, B.~Carro, A.~Sanchez-Esguevillas, and J.~Lloret, ``Network
  traffic classifier with convolutional and recurrent neural networks for
  internet of things,'' \emph{IEEE Access}, vol.~5, pp. 18\,042--18\,050, 2017.

\bibitem{rezaei2018achieve}
S.~Rezaei and X.~Liu, ``How to achieve high classification accuracy with just a
  few labels: A semi-supervised approach using sampled packets,'' \emph{arXiv
  preprint arXiv:1812.09761}, 2018.

\bibitem{kim2018tor}
M.~Kim and A.~Anpalagan, ``Tor traffic classification from raw packet header
  using convolutional neural network,'' in \emph{2018 1st IEEE International
  Conference on Knowledge Innovation and Invention (ICKII)}.\hskip 1em plus
  0.5em minus 0.4em\relax IEEE, 2018, pp. 187--190.

\bibitem{li2011quick}
H.-c. Li, R.~Li, and Q.-h. Liu, ``Quick traffic classification of bt based on
  its handshake packets,'' in \emph{2011 2nd International Conference on
  Artificial Intelligence, Management Science and Electronic Commerce
  (AIMSEC)}.\hskip 1em plus 0.5em minus 0.4em\relax IEEE, 2011, pp. 1336--1339.

\bibitem{shbair2016multi}
W.~M. Shbair, T.~Cholez, J.~Francois, and I.~Chrisment, ``A multi-level
  framework to identify https services,'' in \emph{NOMS 2016-2016 IEEE/IFIP
  Network Operations and Management Symposium}.\hskip 1em plus 0.5em minus
  0.4em\relax IEEE, 2016, pp. 240--248.

\bibitem{crotti2007traffic}
M.~Crotti, M.~Dusi, F.~Gringoli, and L.~Salgarelli, ``Traffic classification
  through simple statistical fingerprinting,'' \emph{ACM SIGCOMM Computer
  Communication Review}, vol.~37, no.~1, pp. 5--16, 2007.

\bibitem{ajaeiya2018mobile}
G.~Ajaeiya, I.~H. Elhajj, A.~Chehab, A.~Kayssi, and M.~Kneppers, ``Mobile apps
  identification based on network flows,'' \emph{Knowledge and Information
  Systems}, vol.~55, no.~3, pp. 771--796, 2018.

\bibitem{li2017traffic}
D.~Li, Y.~Zhu, and W.~Lin, ``Traffic identification of mobile apps based on
  variational autoencoder network,'' in \emph{2017 13th International
  conference on computational intelligence and security (CIS)}.\hskip 1em plus
  0.5em minus 0.4em\relax IEEE, 2017, pp. 287--291.

\bibitem{wang2020real}
X.~Wang, S.~Chen, and J.~Su, ``Real network traffic collection and deep
  learning for mobile app identification,'' \emph{Wireless Communications and
  Mobile Computing}, vol. 2020, 2020.

\bibitem{aceto2017traffic}
G.~Aceto, D.~Ciuonzo, A.~Montieri, and A.~Pescape, ``Traffic classification of
  mobile apps through multi-classification,'' in \emph{GLOBECOM 2017-2017 IEEE
  Global Communications Conference}.\hskip 1em plus 0.5em minus 0.4em\relax
  IEEE, 2017, pp. 1--6.

\bibitem{vinayakumar2019deep}
R.~Vinayakumar, M.~Alazab, K.~Soman, P.~Poornachandran, A.~Al-Nemrat, and
  S.~Venkatraman, ``Deep learning approach for intelligent intrusion detection
  system,'' \emph{IEEE Access}, vol.~7, pp. 41\,525--41\,550, 2019.

\bibitem{shone2018deep}
N.~Shone, T.~N. Ngoc, V.~D. Phai, and Q.~Shi, ``A deep learning approach to
  network intrusion detection,'' \emph{IEEE transactions on emerging topics in
  computational intelligence}, vol.~2, no.~1, pp. 41--50, 2018.

\bibitem{nguyen2019diot}
T.~D. Nguyen, S.~Marchal, M.~Miettinen, H.~Fereidooni, N.~Asokan, and A.-R.
  Sadeghi, ``D{\"i}ot: A federated self-learning anomaly detection system for
  iot,'' in \emph{2019 IEEE 39th International Conference on Distributed
  Computing Systems (ICDCS)}.\hskip 1em plus 0.5em minus 0.4em\relax IEEE,
  2019, pp. 756--767.

\bibitem{rey2021federated}
V.~Rey, P.~M.~S. S{\'a}nchez, A.~H. Celdr{\'a}n, G.~Bovet, and M.~Jaggi,
  ``Federated learning for malware detection in iot devices,'' \emph{arXiv
  preprint arXiv:2104.09994}, 2021.

\bibitem{mclaughlin2017deep}
N.~McLaughlin, J.~Martinez~del Rincon, B.~Kang, S.~Yerima, P.~Miller, S.~Sezer,
  Y.~Safaei, E.~Trickel, Z.~Zhao, A.~Doup{\'e} \emph{et~al.}, ``Deep android
  malware detection,'' in \emph{Proceedings of the Seventh ACM on Conference on
  Data and Application Security and Privacy}, 2017, pp. 301--308.

\bibitem{huang2021network}
X.~Huang, ``Network intrusion detection based on an improved long-short-term
  memory model in combination with multiple spatiotemporal structures,''
  \emph{Wireless Communications and Mobile Computing}, vol. 2021, 2021.

\bibitem{huang2020machine}
H.~Huang, L.~Zhao, H.~Huang, and S.~Guo, ``Machine fault detection for
  intelligent self-driving networks,'' \emph{IEEE Communications Magazine},
  vol.~58, no.~1, pp. 40--46, 2020.

\bibitem{mulvey2018cell}
D.~Mulvey, C.~H. Foh, M.~A. Imran, and R.~Tafazolli, ``Cell coverage
  degradation detection using deep learning techniques,'' in \emph{2018
  International Conference on Information and Communication Technology
  Convergence (ICTC)}.\hskip 1em plus 0.5em minus 0.4em\relax IEEE, 2018, pp.
  441--447.

\bibitem{noshad2019fault}
Z.~Noshad, N.~Javaid, T.~Saba, Z.~Wadud, M.~Q. Saleem, M.~E. Alzahrani, and
  O.~E. Sheta, ``Fault detection in wireless sensor networks through the random
  forest classifier,'' \emph{Sensors}, vol.~19, no.~7, p. 1568, 2019.

\bibitem{rahman2020mockingbird}
M.~S. Rahman, M.~Imani, N.~Mathews, and M.~Wright, ``Mockingbird: Defending
  against deep-learning-based website fingerprinting attacks with adversarial
  traces,'' \emph{IEEE Transactions on Information Forensics and Security},
  vol.~16, pp. 1594--1609, 2020.

\bibitem{attarian2019adawfpa}
R.~Attarian, L.~Abdi, and S.~Hashemi, ``Adawfpa: Adaptive online website
  fingerprinting attack for tor anonymous network: A stream-wise paradigm,''
  \emph{Computer Communications}, vol. 148, pp. 74--85, 2019.

\bibitem{luo2021rbp}
T.~Luo, L.~Wang, S.~Yin, H.~Shentu, and H.~Zhao, ``Rbp: a website
  fingerprinting obfuscation method against intelligent fingerprinting
  attacks,'' \emph{Journal of Cloud Computing}, vol.~10, no.~1, pp. 1--14,
  2021.

\bibitem{7265055}
M.~Conti, L.~V. Mancini, R.~Spolaor, and N.~V. Verde, ``Analyzing android
  encrypted network traffic to identify user actions,'' \emph{IEEE Transactions
  on Information Forensics and Security}, vol.~11, no.~1, pp. 114--125, 2016.

\bibitem{grolman2018transfer}
E.~Grolman, A.~Finkelshtein, R.~Puzis, A.~Shabtai, G.~Celniker, Z.~Katzir, and
  L.~Rosenfeld, ``Transfer learning for user action identication in mobile apps
  via encrypted trafc analysis,'' \emph{IEEE Intelligent Systems}, vol.~33,
  no.~2, pp. 40--53, 2018.

\bibitem{wu2020instagram}
H.~Wu, Q.~Wu, G.~Cheng, and S.~Guo, ``Instagram user behavior identification
  based on multidimensional features,'' in \emph{IEEE INFOCOM 2020-IEEE
  Conference on Computer Communications Workshops (INFOCOM WKSHPS)}.\hskip 1em
  plus 0.5em minus 0.4em\relax IEEE, 2020, pp. 1111--1116.

\bibitem{hou2018classifying}
C.~Hou, J.~Shi, C.~Kang, Z.~Cao, and X.~Gang, ``Classifying user activities in
  the encrypted wechat traffic,'' in \emph{2018 IEEE 37th International
  Performance Computing and Communications Conference (IPCCC)}.\hskip 1em plus
  0.5em minus 0.4em\relax IEEE, 2018, pp. 1--8.

\bibitem{hagos2020machine}
D.~H. Hagos, A.~Yazidi, {\O}.~Kure, and P.~E. Engelstad, ``A machine
  learning-based tool for passive os fingerprinting with tcp variant as a novel
  feature,'' \emph{IEEE Internet of Things Journal}, 2020.

\bibitem{SONG20191}
J.~Song, C.~Cho, and Y.~Won, ``Analysis of operating system identification via
  fingerprinting and machine learning,'' \emph{Computers \& Electrical
  Engineering}, vol.~78, pp. 1--10, 2019.

\bibitem{lastovicka2018passive}
M.~Lastovicka, T.~Jirsik, P.~Celeda, S.~Spacek, and D.~Filakovsky, ``Passive os
  fingerprinting methods in the jungle of wireless networks,'' in \emph{NOMS
  2018-2018 IEEE/IFIP Network Operations and Management Symposium}.\hskip 1em
  plus 0.5em minus 0.4em\relax IEEE, 2018, pp. 1--9.

\bibitem{5440901}
X.~{Zhu}, P.~{Zhang}, X.~{Lin}, and Y.~{Shi}, ``Active learning from stream
  data using optimal weight classifier ensemble,'' \emph{IEEE Transactions on
  Systems, Man, and Cybernetics, Part B (Cybernetics)}, vol.~40, no.~6, pp.
  1607--1621, 2010.

\bibitem{6414645}
I.~{Žliobaitė}, A.~{Bifet}, B.~{Pfahringer}, and G.~{Holmes}, ``Active
  learning with drifting streaming data,'' \emph{IEEE Transactions on Neural
  Networks and Learning Systems}, vol.~25, no.~1, pp. 27--39, 2014.

\bibitem{9012675}
S.~{Wassermann}, T.~{Cuvelier}, P.~{Mulinka}, and P.~{Casas}, ``Adam ral:
  Adaptive memory learning and reinforcement active learning for network
  monitoring,'' in \emph{2019 15th International Conference on Network and
  Service Management (CNSM)}, 2019, pp. 1--9.

\bibitem{settles2008curious}
B.~Settles, ``Curious machines: Active learning with structured instances,''
  Ph.D. dissertation, University of Wisconsin--Madison, 2008.

\bibitem{fundament}
C.~C. Aggarwal, X.~Kong, Q.~Gu, J.~Han, and S.~Y. Philip, ``Active learning: A
  survey,'' in \emph{Data Classification: Algorithms and Applications}.\hskip
  1em plus 0.5em minus 0.4em\relax CRC Press, 2014, pp. 571--605.

\bibitem{konyushkova2017learning}
K.~Konyushkova, R.~Sznitman, and P.~Fua, ``Learning active learning from
  data,'' in \emph{Advances in Neural Information Processing Systems}, 2017,
  pp. 4225--4235.

\bibitem{settles2008analysis}
B.~Settles and M.~Craven, ``An analysis of active learning strategies for
  sequence labeling tasks,'' in \emph{Proceedings of the 2008 Conference on
  Empirical Methods in Natural Language Processing}, 2008, pp. 1070--1079.

\bibitem{settles2011theories}
B.~Settles, ``From theories to queries: Active learning in practice,'' in
  \emph{Active Learning and Experimental Design workshop In conjunction with
  AISTATS 2010}, 2011, pp. 1--18.

\bibitem{shahraki2020boosting}
A.~Shahraki, M.~Abbasi, and {\O}.~Haugen, ``Boosting algorithms for network
  intrusion detection: A comparative evaluation of real adaboost, gentle
  adaboost and modest adaboost,'' \emph{Engineering Applications of Artificial
  Intelligence}, vol.~94, p. 103770, 2020.

\bibitem{yang2011active}
L.~Yang, ``Active learning with a drifting distribution,'' in \emph{Advances in
  Neural Information Processing Systems}, 2011, pp. 2079--2087.

\bibitem{krawczyk2018combining}
B.~Krawczyk, B.~Pfahringer, and M.~Wo{\'z}niak, ``Combining active learning
  with concept drift detection for data stream mining,'' in \emph{2018 IEEE
  International Conference on Big Data (Big Data)}.\hskip 1em plus 0.5em minus
  0.4em\relax IEEE, 2018, pp. 2239--2244.

\bibitem{zhang2018online}
H.~Zhang, W.~Liu, J.~Shan, and Q.~Liu, ``Online active learning paired ensemble
  for concept drift and class imbalance,'' \emph{IEEE Access}, vol.~6, pp.
  73\,815--73\,828, 2018.

\bibitem{costa2018drift}
A.~F.~J. Costa, R.~A.~S. Albuquerque, and E.~M. dos Santos, ``A drift detection
  method based on active learning,'' in \emph{2018 International Joint
  Conference on Neural Networks (IJCNN)}.\hskip 1em plus 0.5em minus
  0.4em\relax IEEE, 2018, pp. 1--8.

\bibitem{torres2019active}
J.~L.~G. Torres, C.~A. Catania, and E.~Veas, ``Active learning approach to
  label network traffic datasets,'' \emph{Journal of information security and
  applications}, vol.~49, p. 102388, 2019.

\bibitem{chen2020malware}
C.-W. Chen, C.-H. Su, K.-W. Lee, and P.-H. Bair, ``Malware family
  classification using active learning by learning,'' in \emph{2020 22nd
  International Conference on Advanced Communication Technology (ICACT)}.\hskip
  1em plus 0.5em minus 0.4em\relax IEEE, 2020, pp. 590--595.

\bibitem{nissim2016aldocx}
N.~Nissim, A.~Cohen, and Y.~Elovici, ``Aldocx: detection of unknown malicious
  microsoft office documents using designated active learning methods based on
  new structural feature extraction methodology,'' \emph{IEEE Transactions on
  Information Forensics and Security}, vol.~12, no.~3, pp. 631--646, 2016.

\bibitem{8058397}
V.~V. {Kumari} and P.~R.~K. {Varma}, ``A semi-supervised intrusion detection
  system using active learning svm and fuzzy c-means clustering,'' in
  \emph{2017 International Conference on I-SMAC (IoT in Social, Mobile,
  Analytics and Cloud) (I-SMAC)}, 2017, pp. 481--485.

\bibitem{springer111}
A.~Beaugnon, P.~Chifflier, and F.~Bach, ``Ilab: An interactive labelling
  strategy for intrusion detection,'' in \emph{Research in Attacks, Intrusions,
  and Defenses}, M.~Dacier, M.~Bailey, M.~Polychronakis, and M.~Antonakakis,
  Eds.\hskip 1em plus 0.5em minus 0.4em\relax Cham: Springer International
  Publishing, 2017, pp. 120--140.

\bibitem{DEKA2019203}
R.~K. Deka, D.~K. Bhattacharyya, and J.~K. Kalita, ``Active learning to detect
  ddos attack using ranked features,'' \emph{Computer Communications}, vol.
  145, pp. 203 -- 222, 2019.

\bibitem{8707963}
Y.~{Zhu} and K.~{Yang}, ``Tripartite active learning for interactive anomaly
  discovery,'' \emph{IEEE Access}, vol.~7, pp. 63\,195--63\,203, 2019.

\bibitem{shu2020generative}
D.~Shu, N.~O. Leslie, C.~A. Kamhoua, and C.~S. Tucker, ``Generative adversarial
  attacks against intrusion detection systems using active learning,'' in
  \emph{Proceedings of the 2nd ACM Workshop on Wireless Security and Machine
  Learning}, 2020, pp. 1--6.

\bibitem{wassermann2019adam}
S.~Wassermann, T.~Cuvelier, P.~Mulinka, and P.~Casas, ``Adam \& ral: Adaptive
  memory learning and reinforcement active learning for network monitoring,''
  in \emph{2019 15th International Conference on Network and Service Management
  (CNSM)}.\hskip 1em plus 0.5em minus 0.4em\relax IEEE, 2019, pp. 1--9.

\bibitem{liu2015active}
S.-M. Liu and Z.-X. Sun, ``Active learning for p2p traffic identification,''
  \emph{Peer-to-Peer Networking and Applications}, vol.~8, no.~5, pp. 733--740,
  2015.

\bibitem{yin2019incorporate}
L.~Yin, H.~Wang, W.~Fan, L.~Kou, T.~Lin, and Y.~Xiao, ``Incorporate active
  learning to semi-supervised industrial fault classification,'' \emph{Journal
  of Process Control}, vol.~78, pp. 88--97, 2019.

\bibitem{yin2018active}
L.~Yin, H.~Wang, and W.~Fan, ``Active learning based support vector data
  description method for robust novelty detection,'' \emph{Knowledge-Based
  Systems}, vol. 153, pp. 40--52, 2018.

\bibitem{dong2021multi}
S.~Dong, ``Multi class svm algorithm with active learning for network traffic
  classification,'' \emph{Expert Systems with Applications}, vol. 176, p.
  114885, 2021.

\bibitem{cheng2013feedback}
Y.~Cheng, Z.~Chen, L.~Liu, J.~Wang, A.~Agrawal, and A.~Choudhary,
  ``Feedback-driven multiclass active learning for data streams,'' in
  \emph{Proceedings of the 22nd ACM international conference on Information \&
  Knowledge Management}, 2013, pp. 1311--1320.

\bibitem{viegas2017toward}
E.~K. Viegas, A.~O. Santin, and L.~S. Oliveira, ``Toward a reliable
  anomaly-based intrusion detection in real-world environments,''
  \emph{Computer Networks}, vol. 127, pp. 200--216, 2017.

\bibitem{draper2016characterization}
G.~Draper-Gil, A.~H. Lashkari, M.~S.~I. Mamun, and A.~A. Ghorbani,
  ``Characterization of encrypted and vpn traffic using time-related,'' in
  \emph{Proceedings of the 2nd international conference on information systems
  security and privacy (ICISSP)}, 2016, pp. 407--414.

\bibitem{lashkari2017characterization}
A.~H. Lashkari, G.~Draper-Gil, M.~S.~I. Mamun, and A.~A. Ghorbani,
  ``Characterization of tor traffic using time based features.'' in
  \emph{ICISSp}, 2017, pp. 253--262.

\bibitem{artstein2008inter}
R.~Artstein and M.~Poesio, ``Inter-coder agreement for computational
  linguistics,'' \emph{Computational Linguistics}, vol.~34, no.~4, pp.
  555--596, 2008.

\bibitem{GARCIA2015108}
L.~P. Garcia, A.~C. {de Carvalho}, and A.~C. Lorena, ``Effect of label noise in
  the complexity of classification problems,'' \emph{Neurocomputing}, vol. 160,
  pp. 108 -- 119, 2015.

\bibitem{perdisci2006misleading}
R.~Perdisci, D.~Dagon, W.~Lee, P.~Fogla, and M.~Sharif, ``Misleading worm
  signature generators using deliberate noise injection,'' in \emph{2006 IEEE
  Symposium on Security and Privacy (S\&P'06)}.\hskip 1em plus 0.5em minus
  0.4em\relax IEEE, 2006, pp. 15--pp.

\bibitem{donmez2008proactive}
P.~Donmez and J.~G. Carbonell, ``Proactive learning: cost-sensitive active
  learning with multiple imperfect oracles,'' in \emph{Proceedings of the 17th
  ACM conference on Information and knowledge management}, 2008, pp. 619--628.

\bibitem{nguyen2004active}
H.~T. Nguyen and A.~Smeulders, ``Active learning using pre-clustering,'' in
  \emph{Proceedings of the twenty-first international conference on Machine
  learning}, 2004, p.~79.

\bibitem{donmez2007dual}
P.~Donmez, J.~G. Carbonell, and P.~N. Bennett, ``Dual strategy active
  learning,'' in \emph{European Conference on Machine Learning}.\hskip 1em plus
  0.5em minus 0.4em\relax Springer, 2007, pp. 116--127.

\bibitem{hakkani2006active}
D.~Hakkani-T{\"u}r, G.~Riccardi, and G.~Tur, ``An active approach to spoken
  language processing,'' \emph{ACM Transactions on Speech and Language
  Processing (TSLP)}, vol.~3, no.~3, pp. 1--31, 2006.

\bibitem{altschuler2019stopping}
M.~Altschuler and M.~Bloodgood, ``Stopping active learning based on predicted
  change of f measure for text classification,'' in \emph{2019 IEEE 13th
  International Conference on Semantic Computing (ICSC)}.\hskip 1em plus 0.5em
  minus 0.4em\relax IEEE, 2019, pp. 47--54.

\bibitem{bloodgood2014method}
M.~Bloodgood and K.~Vijay-Shanker, ``A method for stopping active learning
  based on stabilizing predictions and the need for user-adjustable stopping,''
  \emph{arXiv preprint arXiv:1409.5165}, 2014.

\bibitem{temerinac2015deciding}
M.~Temerinac-Ott, A.~W. Naik, and R.~F. Murphy, ``Deciding when to stop:
  Efficient stopping of active learning guided drug-target prediction,''
  \emph{arXiv preprint arXiv:1504.02406}, 2015.

\bibitem{yang2009outlier}
X.~Yang, L.~J. Latecki, and D.~Pokrajac, ``Outlier detection with globally
  optimal exemplar-based gmm,'' in \emph{Proceedings of the 2009 SIAM
  International Conference on Data Mining}.\hskip 1em plus 0.5em minus
  0.4em\relax SIAM, 2009, pp. 145--154.

\bibitem{ramaswamy2000efficient}
S.~Ramaswamy, R.~Rastogi, and K.~Shim, ``Efficient algorithms for mining
  outliers from large data sets,'' in \emph{Proceedings of the 2000 ACM SIGMOD
  international conference on Management of data}, 2000, pp. 427--438.

\bibitem{breunig2000lof}
M.~M. Breunig, H.-P. Kriegel, R.~T. Ng, and J.~Sander, ``Lof: identifying
  density-based local outliers,'' in \emph{Proceedings of the 2000 ACM SIGMOD
  international conference on Management of data}, 2000, pp. 93--104.

\bibitem{warner2008applied}
R.~M. Warner, \emph{Applied statistics: From bivariate through multivariate
  techniques}.\hskip 1em plus 0.5em minus 0.4em\relax sage, 2008.

\bibitem{vzliobaite2013active}
I.~{\v{Z}}liobait{\.e}, A.~Bifet, B.~Pfahringer, and G.~Holmes, ``Active
  learning with drifting streaming data,'' \emph{IEEE transactions on neural
  networks and learning systems}, vol.~25, no.~1, pp. 27--39, 2013.

\bibitem{dasgupta2005analysis}
S.~Dasgupta, ``Analysis of a greedy active learning strategy,'' in
  \emph{Advances in neural information processing systems}, 2005, pp. 337--344.

\bibitem{freytag2014selecting}
A.~Freytag, E.~Rodner, and J.~Denzler, ``Selecting influential examples: Active
  learning with expected model output changes,'' in \emph{European Conference
  on Computer Vision}.\hskip 1em plus 0.5em minus 0.4em\relax Springer, 2014,
  pp. 562--577.

\bibitem{bouguelia2016adaptive}
M.-R. Bouguelia, Y.~Bela{\"\i}d, and A.~Bela{\"\i}d, ``An adaptive streaming
  active learning strategy based on instance weighting,'' \emph{Pattern
  Recognition Letters}, vol.~70, pp. 38--44, 2016.

\bibitem{abbasi2021deep}
M.~Abbasi, A.~Shahraki, and A.~Taherkordi, ``Deep learning for network traffic
  monitoring and analysis (ntma): A survey,'' \emph{Computer Communications},
  2021.

\bibitem{2020arXiv200900236R}
P.~{Ren}, Y.~{Xiao}, X.~{Chang}, P.-Y. {Huang}, Z.~{Li}, X.~{Chen}, and
  X.~{Wang}, ``{A Survey of Deep Active Learning},'' \emph{arXiv e-prints}, p.
  arXiv:2009.00236, Aug. 2020.

\bibitem{pimentel2018deep}
T.~Pimentel, M.~Monteiro, A.~Veloso, and N.~Ziviani, ``Deep active learning for
  anomaly detection,'' \emph{arXiv preprint arXiv:1805.09411}, 2018.

\bibitem{yang2018active}
K.~Yang, J.~Ren, Y.~Zhu, and W.~Zhang, ``Active learning for wireless iot
  intrusion detection,'' \emph{IEEE Wireless Communications}, vol.~25, no.~6,
  pp. 19--25, 2018.

\bibitem{dai2012peer}
L.~Dai, Y.~Wang, and K.-K. Liu, ``Peer-to-peer traffic identification using
  active learning,'' \emph{Application Research of Computers}, vol.~2, 2012.

\bibitem{yang2019federated}
Q.~Yang, Y.~Liu, T.~Chen, and Y.~Tong, ``Federated machine learning: Concept
  and applications,'' \emph{ACM Transactions on Intelligent Systems and
  Technology (TIST)}, vol.~10, no.~2, pp. 1--19, 2019.

\end{thebibliography}

\begin{IEEEbiography}[{\includegraphics[width=1in,height=1.25in,clip,keepaspectratio]{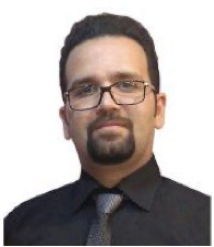}}]{Amin Shahraki}
Amin Shahraki was born in Mashhad, Iran, 1988. He received his bachelor’s degree in Computer Software engineering and master’s degree in Computer Networks in 2009 and 2012 respectively. He received his Ph.D. from University of Oslo in 2020. Now, He is a postdoctoral researcher at School of Computer Science, University College Dublin (UCD), Ireland. He has been a visiting researcher at University of Melbourne, CLOUDS Lab from Aug. 2019 to Feb. 2020 under supervision of Professor Rajkumar Buyya. He is a member of National Elite Foundation of Iran and IEEE since 2011. His current research interests are Internet of Things, Cellular networks, Cognitive Communications and Networking, Network Behavior Analysis, Time Series Analysis, Clustering, QoS and self-healing networks.
\end{IEEEbiography}
\begin{IEEEbiography}[{\includegraphics[width=1in,height=1.25in,clip,keepaspectratio]{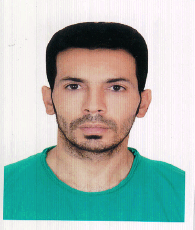}}]{Mahmoud Abbasi}
Mahmoud Abbasi received his M.Sc degree from Department of Computer Engineering, Islamic Azad University of Mashad. He received a B.Eng from the Department of Computer Engineering, Islamic Azad University of Birjand. His current research interests are in the general area of communication systems and networks and \gls{ml}, Internet of Things (IoT), 5G, network traffic analysis, and monitoring (NTMA).
\end{IEEEbiography}
\begin{IEEEbiography}[{\includegraphics[width=1in,height=1.25in,clip,keepaspectratio]{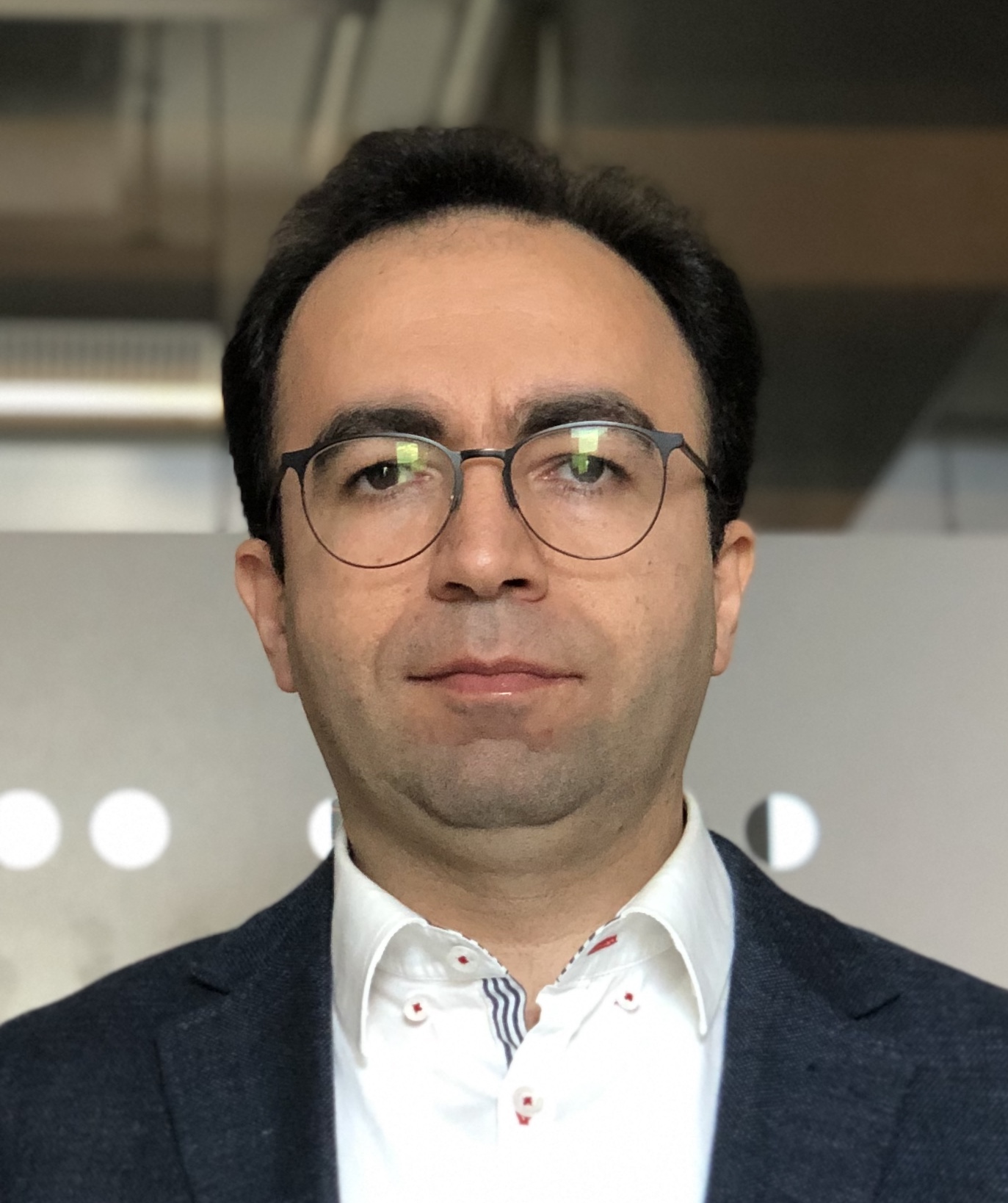}}]{Amir Taherkordi}
is an Associate Professor at the Department of Informatics, University of Oslo (UiO). He received his Ph.D. degree from the Informatics Department, UiO in 2011. After completing his Ph.D. studies, Amir joined Sonitor Technologies as a Senior Embedded Software Engineer. From 2013 to 2018, he was a researcher in the Networks and Distributed Systems (ND) group at the Department of Informatics, UiO. He has so far published several articles in high-ranked con- ferences and journals, and he has experience from several na- tional (Norwegian Research Council) and international (Euro- pean research funding agencies) research projects. Amir’s re- search interests are broadly on resource-efficiency, scalability, adaptability, dependability, mobility and data-intensiveness
\end{IEEEbiography}
\begin{IEEEbiography}[{\includegraphics[width=1in,height=1.25in,clip,keepaspectratio]{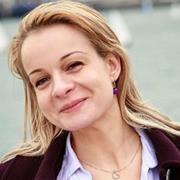}}]{Anca Delia Jurcut} is an Assistant Professor in the UCD School of Computer Science since 2015. She worked as a postdoctoral researcher at University of Limerick and as a Software Engineer in IBM, Dublin in the area of data security and formal verification. Dr. Jurcut research interests include Security Protocols Design and Analysis, Automated Techniques for Formal Verification, Network Security, Attack Detection and Prevention Techniques, Security for the Internet of Things, and Applications of Blockchain for Security and Privacy. Dr.Jurcut has several key contributions in research focusing on detection and prevention techniques of attacks over networks, the design and analysis of security protocols, automated techniques for formal verification, and security for mobile edge computing (MEC). 
 
\end{IEEEbiography}

\end{document}